\documentclass[pss]{wiley2sp} 
\usepackage{amsmath}
\usepackage{bm}              
\usepackage{graphicx}
\usepackage{amssymb}
\usepackage{amsfonts}

\newcommand{\veck}{{\rm\bf k}}
\newcommand{\vecp}{{\rm\bf p}}
\newcommand{\vecR}{{\rm\bf R}}
\newcommand{\vecS}{{\rm\bf S}}
\newcommand{\vecRprime}{{\rm\bf R}^{\prime}}
\newcommand{\vecQ}{{\rm\bf Q}}
\newcommand{\vecG}{{\rm\bf G}}
\newcommand{\veczero}{{\rm\bf 0}}
\newcommand{\rmi}{{\rm i}}
\newcommand{\rmd}{{\rm d}}
\newcommand{\openone}{\leavevmode\hbox{\small1\normalsize\kern-.33em1}}
\renewcommand{\det}{\Delta}

\newcommand{\mytoday}{Version of December 16, 2016, to be submitted}

\def\Xint#1{\mathchoice
   {\XXint\displaystyle\textstyle{#1}}%
   {\XXint\textstyle\scriptstyle{#1}}%
   {\XXint\scriptstyle\scriptscriptstyle{#1}}%
   {\XXint\scriptscriptstyle\scriptscriptstyle{#1}}%
   \!\int}
\def\XXint#1#2#3{{\setbox0=\hbox{$#1{#2#3}{\int}$}
     \vcenter{\hbox{$#2#3$}}\kern-.5\wd0}}

\def\dashint{\Xint-}

\begin{document}

\title{Non-interacting two-impurity\\ Anderson model on a lattice\\ 
at particle-hole symmetry}

\titlerunning{Non-interacting TIAM on a lattice}

\author{%
Zakaria M.M.\ Mahmoud\textsuperscript{\textsf{\bfseries 1,2}},
J\"org B\"unemann\textsuperscript{\textsf{\bfseries 3}},
Florian Gebhard\textsuperscript{\textsf{\Ast,\bfseries 1}}}

\authorrunning{Z.M.M.\ Mahmoud et al.}

\mail{e-mail \textsf{florian.gebhard@physik.uni-marburg.de}, Phone:
  +49-6421-2821318, Fax: +49-6421-2824511}

\institute{\textsuperscript{1}\,Fachbereich Physik, Philipps-Universit\"at Marburg,
D-35032 Marburg, Germany\\
\textsuperscript{2}\,Department of Physics, New-Valley Faculty of Science, 
El-Kharga, Assiut University, Egypt\\
\textsuperscript{3}\,Institut f\"ur Physik, BTU Cottbus-Senftenberg, D-03013 Cottbus, 
Germany}

\received{XXXX, revised XXXX, accepted XXXX} 
\published{XXXX} 

\keywords{Impurity scattering, RKKY interaction.}

\date{\mytoday}

\abstract{
\vspace*{5pt}
\abstcol{%
We study the non-interacting two-impurity Anderson model on a lattice 
using the Green function equation-of-motion method. 
A case of particular interest is the RKKY limit 
that is characterized by a small hybridization between
impurities and host electrons and the absence 
of a direct coupling between the impurities.
In contrast to the low-density case, at half band-filling and particle-hole symmetry, 
the RKKY interaction decays as}%
{the inverse square of the impurity distance 
along the axis of a simple cubic lattice. In the RKKY limit, for the spectral function 
we generically observe a small splitting of the single-impurity 
resonance into two peaks. For a vanishing density-density correlation function
of the host electrons, we find only a broadened single peak in the local density of 
states.}\vspace*{5pt}}
\maketitle

\section{Introduction}

Impurities diluted in a metallic host 
pose a fundamental problem in solid-state theory~\cite{Solyom}.
A famous example is the formation of a `Kondo cloud'
around a magnetic impurity in a metallic host that leads to a complete screening
of a spin-1/2 impurity at zero temperature and to a very narrow
Abrikosov-Suhl resonance in the single-particle density of states~\cite{Hewson}.
The formation of a magnetic impurity and its screening by
the host electrons is contained in the single-impurity
Anderson model~\cite{PhysRev.124.41}; for a review, see Ref.~\cite{Hewson}, 
and references therein.

A single (magnetic) impurity induces 
distortions in the host electrons' charge and spin density
(`Friedel oscillations')~\cite{Fridelphilmag,Friedeladvances}.
These distortions can be sensed by a second (magnetic) 
impurity so that the host electrons generate
an effective interaction between the impurities, known as
RKKY interaction, named after Ruderman and Kittel~\cite{PhysRev.96.99},
Kasuya~\cite{Kasuya01071956}, and Yosida~\cite{PhysRev.106.893};
for a concise derivation from perturbation theory, see appendix~I of~\cite{Solyom}.

The combined physics of the Kondo effect and of the RKKY interaction is contained
in the two-impurity Anderson model (TIAM)~\cite{AlexanderAnderson}.
It describes two impurities embedded in a metallic host at lattice sites
$\vecR_1$ and $\vecR_2$. When the local Hubbard interaction on the impurities
is strong, the model covers the local Kondo physics
and the RKKY interaction between magnetic impurities.
However, the TIAM
is a true many-particle problem that cannot be solved
in general; for a recent investigation using an extended non-crossing approximation, 
see Ref.~\cite{Grewe}, and references therein.

The TIAM is frequently invoked in studies of coupled quantum 
dots where each dot represents an impurity,
see Ref.~\cite{quantumdots} for a recent study. However, the quantum dots
have individual leads, i.e., there are two independent host metals
so that the indirect exchange interaction is different from the solid-state case
for which the TIAM was designed originally. Moreover, 
the direct coupling between the quantum dots is generically large and dominates over
the RKKY interaction.

In the present work, we consider the non-interacting 
two-impurity Anderson model
that can be solved exactly using the Green function equation-of-motion 
method~\cite{AlexanderAnderson}. 
In contrast to the perturbative RKKY derivation, the results include the 
full impurity-host hybridization and multiple scattering events of the host electrons
off the impurities to all orders.
The formulae provide the basis of a Gutzwiller
approach to the TIAM which permits a variational
analysis of the competition between the single-impurity Kondo effect and the 
two-impurity RKKY interaction~\cite{GutzwillerTIAM}.
For this reason, we are particularly interested 
in the case of particle-hole symmetry at half band-filling.

To study the competition between single-impurity and two-impurity physics,
we focus on the RKKY limit of a small, local hybridization
in the absence of a direct electron transfer between the impurities.
This case was not worked out in detail in Ref.~\cite{AlexanderAnderson}
where the impurities were dominantly coupled by a direct electron transfer.
Although we cannot study the formation of a local moments,
we determine the effective RKKY interaction of the (non-interacting) impurities as
a function of their separation fully analytically.
We find that the RKKY interaction generically leads to two peaks 
in the single-particle density of states, i.e., the host metal generates
a small but finite transfer matrix element between the impurities.

Our work is organized as follows.
We formulate the two-impurity Anderson model in section~\ref{sec:TIAMdef}.
In  section~\ref{sec:nonintmodel} we solve the non-interacting model 
using the equation-of-motion method for the retarded Green 
functions~\cite{AlexanderAnderson}.
We introduce and utilize particle-hole symmetry
in section~\ref{sec:phsymm}, and work out the impurity properties 
for tight-binding host electrons on a simple-cubic lattice
in detail in section~\ref{sec:TBmodel}.
Short conclusions, section~\ref{sec:conclusions}, end our presentation. 
Some technical details are deferred to the appendix.

\section{Two-impurity Anderson model}
\label{sec:TIAMdef}

We start our investigations with the definition of the Hamiltonian. 
Then, we rephrase the problem in terms of a single-site two-orbital model.

\subsection{Hamiltonian}

Two impurities in a metallic host on a lattice 
are modeled by the Hamiltonian~\cite{AlexanderAnderson}
\begin{equation}
\hat{H}=\hat{T}+\hat{T}_d+\hat{H}_{\rm int}+\hat{V} \; .
\label{eq:defH}
\end{equation}
Here, $\hat{T}$ is the kinetic energy of the non-interacting spin-1/2
host electrons ($\sigma=\uparrow,\downarrow$), 
\begin{equation}
\hat{T}=  \sum_{\vecR,\vecRprime;\sigma}t(\vecR-\vecRprime)
\hat{c}_{\vecR,\sigma}^+\hat{c}_{\vecRprime,\sigma}^{\vphantom{+}}\; ,
\label{eq:defT}
\end{equation}
where the electrons tunnel between the sites $\vecR$ and $\vecRprime$
of the lattice with amplitude $t(\vecR-\vecRprime)$.
The kinetic energy is diagonal in Fourier space.
For $\veck$ from the first Brillouin zone we define
\begin{equation}
\hat{c}_{\veck,\sigma}^{\vphantom{+}}=
\sqrt{\frac{1}{L}} \sum_{\vecR} e^{-\rmi \veck \cdot \vecR}
\hat{c}_{\vecR,\sigma}^{\vphantom{+}}\, , \, 
\hat{c}_{\vecR,\sigma}^{\vphantom{+}}=
\sqrt{\frac{1}{L}} \sum_{\vecR} e^{\rmi \veck \cdot \vecR}
\hat{c}_{\veck,\sigma}^{\vphantom{+}}\, ,
\end{equation}
where $L$ is the (even) number of lattice sites. With
\begin{equation}
t(\vecR) = \frac{1}{L} \sum_{\veck} e^{\rmi \veck\cdot \vecR} \epsilon(\veck)
\; , \; \epsilon(\veck)=\sum_{\vecR}t(\vecR)e^{-\rmi \veck\cdot \vecR}
\end{equation}
we find
\begin{equation}
\hat{T}= \sum_{\veck,\sigma} \epsilon(\veck) 
\hat{c}_{\veck,\sigma}^+\hat{c}_{\veck,\sigma}^{\vphantom{+}} \;,
\end{equation}
where $\epsilon(\veck)$ is the dispersion relation.

With $\hat{T}_d$ we also permit a direct electron transfer 
with amplitude $t_{12}$ between the impurity orbitals at sites $\vecR_1$ and $\vecR_2$,
\begin{equation}
\hat{T}_{d}=
\sum_{\sigma}t_{12}\hat{d}_{1,\sigma}^+\hat{d}_{2,\sigma}^{\vphantom{+}}
+ t_{12}^*\hat{d}_{2,\sigma}^+\hat{d}_{1,\sigma}^{\vphantom{+}}
\; .
\label{eq:defTd}
\end{equation}

Next, $\hat{H}_{\rm int}$ represents the Hubbard interaction
to model the Coulomb repulsion on the impurities,
\begin{equation}
\hat{H}_{\rm int}= U \sum_{b=1}^2 
(\hat{n}_{b,\uparrow}-1/2)(\hat{n}_{b,\downarrow}-1/2)\; ,
\label{eq:defHint}
\end{equation}
where $\hat{n}_{b,\sigma}=\hat{d}_{b,\sigma}^+\hat{d}_{b,\sigma}^{\vphantom{+}}$
counts the number of impurity electrons ($b=1,2$).

Lastly, $\hat{V}$ describes the hybridization between impurity and host electron
states,
\begin{eqnarray}
\hat{V}&=&  
\sum_{\vecR,b,\sigma}
V(\vecR-\vecR_b)
\hat{c}_{\vecR,\sigma}^+\hat{d}_{b,\sigma}^{\vphantom{+}}
+ V^*(\vecR-\vecR_b) \hat{d}_{b,\sigma}^+\hat{c}_{\vecR,\sigma}^{\vphantom{+}}
\nonumber \\
&=&
\sqrt{\frac{1}{L}}\sum_{\veck,b,\sigma}
V_{\veck}e^{-\rmi \veck \cdot \vecR_b}
\hat{c}_{\veck,\sigma}^+\hat{d}_{b,\sigma}^{\vphantom{+}}
+ V_{\veck}^* e^{\rmi \veck\cdot \vecR_b} 
\hat{d}_{b,\sigma}^+\hat{c}_{\veck,\sigma}^{\vphantom{+}}\; ,\nonumber \\
V_{\veck}&=& \sum_{\vecR}V(\vecR)e^{-\rmi \veck\cdot \vecR} \,,\,
V(\vecR)=\frac{1}{L}\sum_{\veck}e^{\rmi \veck \cdot \vecR}V_{\veck} \; .
\label{eq:defV}
\end{eqnarray}
The model~(\ref{eq:defH}) poses a difficult many-particle problem
that cannot be solved in general.

\subsection{Single-site two-orbital model}

As a second step,
we map the two-impurity model onto an asymmetric two-orbital model.

\subsubsection{Kinetic energy of d-electrons}

We introduce the $h$-basis for the impurity electrons using the unitary transformation
\begin{eqnarray}
\hat{d}_{1,\sigma}^+&=& \sqrt{\frac{1}{2}} \left( \hat{h}_{1,\sigma}^+
+\alpha_{12} \hat{h}_{2,\sigma}^+\right)  \; ,\nonumber \\
\hat{d}_{2,\sigma}^+&=& \sqrt{\frac{1}{2}} \left(-\alpha_{12}^* \hat{h}_{1,\sigma}^+
+\hat{h}_{2,\sigma}^+\right)  \; ,
\end{eqnarray}
where 
\begin{equation}
\alpha_{12}=\frac{t_{12}^*}{|t_{12}|} \quad, \quad  
\alpha_{12}^2=\frac{t_{12}^*}{t_{12}}
\;.
\label{eq:defalpha12}
\end{equation}
The inverse transformation reads
\begin{eqnarray}
\hat{h}_{1,\sigma}^+&=& \sqrt{\frac{1}{2}} \left( \hat{d}_{1,\sigma}^+
-\alpha_{12} \hat{d}_{2,\sigma}^+\right) \nonumber \; , \\
\hat{h}_{2,\sigma}^+&=& \sqrt{\frac{1}{2}} \left( \alpha_{12}^* \hat{d}_{1,\sigma}^+
+\hat{d}_{2,\sigma}^+\right) \nonumber \; . 
\end{eqnarray}
Then, $\hat{T}_d$, eq.~(\ref{eq:defTd}), is diagonal in the $h$-basis,
\begin{equation}
\hat{T}_d= |t_{12}| \left( 
\hat{h}_{2,\sigma}^+\hat{h}_{2,\sigma}^{\vphantom{+}}
-\hat{h}_{1,\sigma}^+\hat{h}_{1,\sigma}^{\vphantom{+}} 
\right) \; .
\end{equation}
In this representation, $\hat{T}_d$ has the form of a splitting 
of two impurity levels on the same site.

For a unitary transformation we have
\begin{equation}
\hat{d}_{1,\sigma}^+\hat{d}_{1,\sigma}^{\vphantom{+}} 
+ 
\hat{d}_{2,\sigma}^+\hat{d}_{2,\sigma}^{\vphantom{+}} 
= 
\hat{h}_{1,\sigma}^+\hat{h}_{1,\sigma}^{\vphantom{+}} 
+
\hat{h}_{2,\sigma}^+\hat{h}_{2,\sigma}^{\vphantom{+}}  \; .
\end{equation}
Therefore, the average number of $d_{\sigma}$-electrons equals the 
average number of $h_{\sigma}$-electrons.

\subsubsection{Hybridization}

In the $h$-basis, the hybridization $\hat{V}$, see eq.~(\ref{eq:defV}),
takes the form
\begin{equation}
\hat{V}=  \sqrt{\frac{1}{L}}\sum_{\veck,b,\sigma}
V_{\veck,b}\hat{c}_{\veck,\sigma}^+\hat{h}_{b,\sigma}^{\vphantom{+}}
+ V_{\veck,b}^* 
\hat{h}_{b,\sigma}^+\hat{c}_{\veck,\sigma}^{\vphantom{+}}\; .
\label{eq:defVdiagonal}
\end{equation}
The two impurity levels hybridize with the conduction electrons with 
the matrix elements
\begin{eqnarray}
V_{\veck,1}&=&\frac{V_{\veck}}{\sqrt{2}}
\left( e^{-\rmi \veck\cdot \vecR_1}
-\alpha_{12} e^{-\rmi \veck\cdot \vecR_2}\right) \; , \nonumber \\
V_{\veck,2}&=&\frac{V_{\veck}}{\sqrt{2}}
\left( \alpha_{12}^* e^{-\rmi \veck\cdot \vecR_1}
+ e^{-\rmi \veck\cdot \vecR_2}\right) \; .
\label{eq:hybridizationsforh}
\end{eqnarray}
We do not elaborate $\hat{H}_{\rm int}$ in the new basis because
we restrict ourselves to the non-interacting model in the following.

\section{Green functions for the non-interacting model}
\label{sec:nonintmodel}
For $\, U=0$ in~(\ref{eq:defH}),
the single-particle Green functions can be calculated exactly using the 
equation-of-motion method. From the Green functions, all ground-state properties
and the single-particle density of states can be derived.

\subsection{Definition}

The retarded Green functions for the host electrons and the impurity
electrons read
\begin{eqnarray}
G_{\veck,\vecp;\sigma}(t) &=& (-\rmi) \Theta(t) \langle \Phi_0 |
\bigl[ \hat{c}_{\veck,\sigma}^{\vphantom{+}}(t) ,
\hat{c}_{\vecp,\sigma}^+\bigr]_+ | \Phi_0\rangle \; , 
\label{eq:Gkp}\\
G_{b,b';\sigma}(t) &=& (-\rmi) \Theta(t) \langle \Phi_0 |
\bigl[ \hat{h}_{ b,\sigma}^{\vphantom{+}}(t) ,
\hat{h}_{b',\sigma}^+\bigr]_+ | \Phi_0\rangle \; ,
\label{eq:Gbbprime}
\end{eqnarray}
where $|\Phi_0\rangle$ is the ground state for $U=0$,
$\Theta(t)$ is the Heaviside step function, 
$[\hat{A},\hat{B}]_+=\hat{A}\hat{B}+\hat{B}\hat{A}$
is the anti-commutator, and
\begin{equation}
\hat{A}(t) = e^{\rmi \hat{H} t} \hat{A} e^{-\rmi \hat{H} t} 
\end{equation}
describes the time evolution of a Schr\"odinger operator $\hat{A}$
in the Heisenberg picture; we set $\hbar\equiv 1$
for convenience.

The host and impurity Green 
functions in eqs.~(\ref{eq:Gkp}) and~(\ref{eq:Gbbprime}) 
couple to the mixed Green functions
\begin{eqnarray}
G_{b,\veck;\sigma}(t) &=& (-\rmi) \Theta(t) \langle \Phi_0 |
\bigl[ \hat{h}_{ b,\sigma}^{\vphantom{+}}(t) ,
\hat{c}_{\veck,\sigma}^+\bigr]_+ | \Phi_0\rangle \; ,
\label{eq:Gbk}\\
G_{\veck,b;\sigma}(t) &=& (-\rmi) \Theta(t) \langle \Phi_0 |
\bigl[ \hat{c}_{\veck,\sigma}^{\vphantom{+}}(t) ,
\hat{h}_{b,\sigma}^+\bigr]_+ | \Phi_0\rangle \; ,
\label{eq:Gkb}
\end{eqnarray}
when we set up the equations of motion.

\subsection{Equations of motion}

The time derivatives of the single-electron Heisenberg operators read
\begin{eqnarray}
\rmi \frac{\rmd}{\rmd t}
\hat{c}_{\veck,\sigma}^{\vphantom{+}}(t) 
&=& e^{\rmi \hat{H} t} \left[\hat{c}_{\veck,\sigma}^{\vphantom{+}}(t), \hat{H}
\right]_{-} e^{-\rmi \hat{H} t} \nonumber \\
&=& \epsilon(\veck) \hat{c}_{\veck,\sigma}^{\vphantom{+}}(t) 
+\sqrt{\frac{1}{L}}\sum_b V_{\veck,b} \hat{h}_{b,\sigma}^{\vphantom{+}}(t) \; ,\\
\rmi \frac{\rmd}{\rmd t}
\hat{h}_{1,\sigma}^{\vphantom{+}}(t) 
&=& -|t_{12}| \hat{h}_{1,\sigma}^{\vphantom{+}}(t) 
+\sqrt{\frac{1}{L}}
\sum_{\veck} V_{\veck,1}^* \hat{c}_{\veck,\sigma}^{\vphantom{+}}(t) \; ,\\
\rmi \frac{\rmd}{\rmd t}
\hat{h}_{2,\sigma}^{\vphantom{+}}(t) 
&=& |t_{12}| \hat{h}_{2,\sigma}^{\vphantom{+}}(t) 
+\sqrt{\frac{1}{L}}\sum_{\veck} V_{\veck,2}^* \hat{c}_{\veck,\sigma}^{\vphantom{+}}(t) \; .
\end{eqnarray}
Moreover, we define the Fourier transformation 
for a retarded Green function $g(t)$ ($\omega\equiv 
\omega+\rmi \eta$, $\eta=0^+$)
\begin{equation}
g(\omega) = \int_0^{\infty} \rmd t e^{\rmi \omega t} g(t)
\; ,\; 
g(t) = \int_{-\infty}^{\infty} \frac{\rmd \omega}{2\pi} 
e^{-\rmi \omega t} g(\omega) \; .
\label{eq:FourierT}
\end{equation}
With these equations it is possible to derive a closed set
of algebraic equations for the Green functions.
For later use, we define the bare Green functions
\begin{eqnarray}
g_{\veck}(\omega) &=& \frac{1}{\omega-\epsilon(\veck)} \; ,
\nonumber \\
g_1(\omega) &=& \frac{1}{\omega+|t_{12}|} \; , \nonumber \\
g_2(\omega) &=& \frac{1}{\omega-|t_{12}|} 
\label{eq:defgb}
\end{eqnarray}
that appear in the absence of the hybridization, $V_{\veck}=0$.

\subsubsection{Host electrons}

The first set of equations involves the host electrons,
\begin{eqnarray}
\rmi \dot{G}_{\veck,\vecp}(t) &=& \delta(t) \delta_{\veck,\vecp}
+ \epsilon(\veck) G_{\veck,\vecp}(t)\nonumber \\
&& + \sqrt{\frac{1}{L}}
\sum_b V_{\veck,b} G_{b,\vecp}(t)  \; ,\nonumber \\
\rmi \dot{G}_{1,\vecp}(t) &=& 
-|t_{12}| G_{1,\vecp}(t)
+ \sqrt{\frac{1}{L}}\sum_{\veck} V_{\veck,1}^* G_{\veck,\vecp}(t)  \; ,\nonumber \\
\rmi \dot{G}_{2,\vecp}(t) &=& 
|t_{12}| G_{2,\vecp}(t)
+ \sqrt{\frac{1}{L}}\sum_{\veck} V_{\veck,2}^* G_{\veck,\vecp}(t)  \; ,
\end{eqnarray}
where we dropped the spin index for convenience.
After Fourier transformation, we find that
\begin{equation}
G_{b,\vecp}(\omega) = A_{b,\vecp}(\omega)g_b(\omega)
\end{equation}
with 
\begin{equation}
A_{b,\vecp}(\omega)= \sqrt{\frac{1}{L}}
\sum_{\veck} V_{\veck,b}^* G_{\veck,\vecp}(\omega)
\label{eq:definesAb}
\end{equation}
so that the host Green function obeys 
\begin{equation}
  G_{\veck,\vecp}(\omega) = g_{\veck}(\omega)
\Bigl( \delta_{\veck,\vecp} +\sqrt{\frac{1}{L}}
\sum_b V_{\veck,b}A_{b,\vecp}(\omega)g_b(\omega)\Bigr)\; .
\end{equation}
When we insert this expression into eq.~(\ref{eq:definesAb})
we find that
\begin{eqnarray}
A_{b,\vecp}(\omega)&=& \sqrt{\frac{1}{L}}V_{\vecp,b}^* g_{\vecp}(\omega)
+ A_{b,\vecp}(\omega)g_{b,\sigma}(\omega) H_{b,b}(\omega) \nonumber \\
&& + A_{\bar{b},\vecp}(\omega)g_{\bar{b}}(\omega) H_{\bar{b},b}(\omega) \; ,
\label{eq:defAbpsigma}
\end{eqnarray}
where we introduced the notation $\bar{1}=2$ and $\bar{2}=1$, and the 
hybridization matrix functions
\begin{equation}
H_{b,b'}(\omega) = \frac{1}{L}\sum_{\veck}
\frac{V_{\veck,b}V_{\veck,b'}^*}{\omega-\epsilon(\veck)} \; .
\label{eq:defHbbprime}
\end{equation}
The resulting $2\times 2$ matrix problem~(\ref{eq:defAbpsigma}) 
for fixed frequency~$\omega$ and fixed Bloch momentum~$\vecp$
is readily solved to give
\begin{equation}
g_b(\omega) 
A_{b,\vecp}(\omega) =\frac{Z_{b,\vecp}(\omega)}{\det(\omega)} \; ,
\end{equation}
where
\begin{eqnarray}
Z_{b,\vecp}(\omega) &=& 
\sqrt{\frac{1}{L}}V_{\vecp,b}^*g_{\vecp}(\omega)
\left(g_{\bar{b}}^{-1}(\omega)-H_{\bar{b},\bar{b}}(\omega)\right)\nonumber \\
&& + \sqrt{\frac{1}{L}}
V_{\vecp,\bar{b}}^*g_{\vecp}(\omega) H_{\bar{b},b}(\omega)\; ,\nonumber \\
\det(\omega)&=& 
\left(g_2^{-1}(\omega)-H_{2,2}(\omega)\right)
\left(g_1^{-1}(\omega)-H_{1,1}(\omega)\right) \nonumber \\
&&- H_{1,2}(\omega)H_{2,1}(\omega) \; .
\label{eq:resultGBk}
\end{eqnarray}
The trace over all $\veck$-states results in the average host Green function
\begin{equation}
\sum_{\veck} G_{\veck,\veck}(\omega) =
G_{\sigma,0}(\omega) + \Delta G_{\sigma}^{\rm host}(\omega)
\end{equation}
with
\begin{eqnarray}
G_{\sigma,0}(\omega)&=& \sum_{\veck} g_{\veck}(\omega) \; , \nonumber \\
\Delta G_{\sigma}^{\rm host}(\omega) &=&
\sum_{\veck,b} 
\frac{[g_{\veck}(\omega)]^2 V_{\veck,b}V_{\veck,b}^*}{L\det(\omega)} 
\bigl(g_{\bar{b}}^{-1}(\omega) -H_{\bar{b},\bar{b}}(\omega)\bigr)  
\nonumber \\
&& \hphantom{\sum_{\veck,b} }
+ \frac{[g_{\veck}(\omega)]^2 V_{\veck,b}V_{\veck,\bar{b}}^*}{L\det(\omega)} 
H_{\bar{b},b}(\omega) 
 \nonumber \\
&=& \sum_b \frac{g_{\bar{b}}^{-1}(\omega) -H_{\bar{b},\bar{b}}(\omega)}{
\det(\omega)} \bigl(-H'_{b,b}(\omega)\bigr)
\nonumber \\
&& \hphantom{\sum_b} 
+ \frac{H_{\bar{b},b}(\omega)}{\det(\omega)} 
\bigl( -H'_{b,\bar{b}}(\omega)\bigr)
\label{eq:tracek}
\end{eqnarray}
with $H'_{b,b'}(\omega)=(\partial H_{b,b'}(\omega))/(\partial \omega)$.
We need the average host Green function
for the calculation of the ground-state energy and
the impurity contribution to the density of states.

\subsubsection{Impurity electrons}

After Fourier transformation
the equations of motion for the impurity Green functions read 
\begin{eqnarray}
g_b^{-1}(\omega) G_{b,b'}(\omega) &=& \delta_{b,b'} 
+ \sqrt{\frac{1}{L}}\sum_{\veck}V_{\veck,b}^*G_{\veck,b'}(\omega) \; ,\nonumber \\
g_{\veck}^{-1}(\omega) G_{\veck,b}(\omega) &=& \sqrt{\frac{1}{L}}
\sum_{b'}V_{\veck,b'}G_{b',b}(\omega)
\; .
\end{eqnarray}
We insert the second equation into the first to obtain
\begin{equation}
g_b^{-1}(\omega) G_{b,b'}(\omega) = \delta_{b,b'} 
+ \sum_{b''} H_{b'',b}(\omega)G_{b'',b'}(\omega) \; .
\end{equation}
The solution of this $2\times2$ matrix problem for fixed $\omega$ gives
\begin{equation}
G_{b,\bar{b}}(\omega)= \frac{H_{\bar{b},b}(\omega)}{\det(\omega)} \; ,
\end{equation}
and 
\begin{equation}
G_{b,b}(\omega)= \frac{g_{\bar{b}}^{-1}(\omega)-H_{\bar{b},\bar{b}}(\omega)}{
\det(\omega)} \; .
\label{eq:Gbbdiagonal}
\end{equation}
For our further investigations, 
the diagonal impurity Green functions in eq.~(\ref{eq:Gbbdiagonal})
are sufficient.

\subsection{Ground-state expectation values}

Lastly, we use the Green functions to calculate the impurity
contribution to the single-particle density of states. In addition, we
derive the ground-state energy, impurity density, and hybridization energy.

\subsubsection{Density of states and ground-state energy}

We introduce the single-particle density of states 
\begin{equation}
D_{\sigma}(\omega) = \sum_{\alpha} \delta(\omega-E_{\alpha}) \; , 
\end{equation}
where $E_{\alpha}$ are the energies of the single-particle
levels. We introduce 
the exact single-particle and single-hole excitations of the ground state
$|\alpha\rangle \equiv \hat{a}_{\alpha,\sigma}^{(+)}| \Phi_0 \rangle$,
\begin{equation}
\hat{H} \hat{a}_{\alpha,\sigma}^{(+)}| \Phi_0 \rangle 
= (E_0\pm E_{\alpha}) \hat{a}_{\alpha,\sigma}^{(+)}| \Phi_0 \rangle \; ,
\end{equation}
where $E_0$ is the energy of the ground state $|\Phi_0\rangle$.
Then, the single-particle density of states can be written as
\begin{eqnarray}
D_{\sigma}(\omega)&=& - \frac{1}{\pi} {\rm Im}\biggl[ 
\langle \Phi_0 | \hat{a}_{\alpha,\sigma}^+ 
\frac{1}{\omega-(\hat{H}-E_0)+\rmi \eta} \hat{a}_{\alpha,\sigma}^{\vphantom{+}} 
| \Phi_0 \rangle \nonumber \\
&& \hphantom{\frac{1}{\pi} }
+ \langle \Phi_0 | \hat{a}_{\alpha,\sigma}^{\vphantom{+}} 
\frac{1}{\omega-(\hat{H}-E_0)+\rmi \eta} \hat{a}_{\alpha,\sigma}^+
| \Phi_0 \rangle \biggr]\, .\nonumber \\
\end{eqnarray}
The sum on all single-particle excitations is equivalent to the trace
over the subspace of all single-particle excitations,
\begin{equation}
D_{\sigma}(\omega)= - \frac{1}{\pi} {\rm Im}
{\rm Tr}_1 \left[\frac{1}{\omega-(\hat{H}-E_0)+\rmi \eta} \right] \; .
\end{equation}
We can equally use the excitations
$\hat{c}_{\veck,\sigma}^{(+)}| \Phi_0 \rangle$ 
and 
$\hat{h}_{b,\sigma}^{(+)}| \Phi_0 \rangle$ 
to perform the trace over the single-particle excitations
of the ground state. Therefore, we may write
\begin{eqnarray}
D_{\sigma}(\omega)
& =& - \frac{1}{\pi} {\rm Im}\left[ 
\sum_{\veck} G_{\veck,\veck}(\omega) +\sum_b G_{b,b}(\omega)
\right] \nonumber \\
&=& D_{\sigma,0}(\omega) +
\Delta D_{\sigma}^{\rm host}(\omega) +\sum_b D_{b,\sigma}(\omega)
\end{eqnarray}
with
\begin{equation}
D_{\sigma,0}(\omega)=\sum_{\veck}\delta\bigl(\omega-\epsilon(\veck)\bigr)
=-\frac{1}{\pi} \sum_{\veck}{\rm Im}\left[g_{\veck}(\omega)\right]\; .
\end{equation}
Moreover,
\begin{eqnarray}
D_{b,\sigma}(\omega) &=& -\frac{1}{\pi} {\rm Im}
\left[ G_{b,b}(\omega)\right] \nonumber \\
&=& -\frac{1}{\pi} {\rm Im}\biggl[
\frac{g_{\bar{b}}^{-1}(\omega)-H_{\bar{b},\bar{b}}(\omega)}{\det(\omega)}
\biggr] \nonumber \; ,\\
\Delta D_{\sigma}^{\rm host}(\omega) &=& -\frac{1}{\pi} {\rm Im}
\left[ \Delta G_{\sigma}^{\rm host}(\omega)\right] 
\end{eqnarray}
with $g_b(\omega)$ from eq.~(\ref{eq:defgb}), $H_{b,b'}(\omega)$ from
eq.~(\ref{eq:defHbbprime}), and
$\Delta G_{\sigma}^{\rm host}(\omega)$ from eq.~(\ref{eq:tracek}).
Therefore, we find
\begin{eqnarray}
D_{{\rm imp},\sigma}(\omega)&=& D_{\sigma}(\omega)-D_{\sigma,0}(\omega)
\nonumber \\
&=& \Delta D_{\sigma}^{\rm host}(\omega) + \sum_b D_{b,\sigma}(\omega) 
\end{eqnarray}
for the contribution to the density of states that originates from
the impurities and their hybridization to the host electrons.

The change in the ground-state energy contribution due to the hybridization 
between host electrons and impurities is given by
\begin{equation}
\Delta E_0 = 2 \sum_{\alpha}^{\rm occ} \left(E_{\alpha}-\epsilon_{\alpha}\right)
\; ,
\end{equation}
where $E_{\alpha}$ ($\epsilon_{\alpha}$) are the single-particle energies 
in the presence (absence) of the hybridization $V_{\veck}$,
and the factor two accounts for the spin degeneracy.
With the help of the density of states it is readily calculated from
\begin{equation}
\Delta E_0 = 2 \biggl[ |t_{12}| 
+\int_{-\infty}^0 \rmd \omega \omega D_{{\rm imp},\sigma}(\omega) \biggr]
\; .
\label{eq:gsenergyfromDOS}
\end{equation}

\subsubsection{Particle density and hybridization energy}

In general, for the retarded Green function
\begin{equation}
G_{A,B}(t)= (-\rmi) \Theta(t) \langle \Phi_0 |
\left[ \hat{A}^{\vphantom{+}}(t) ,
\hat{B}^+\right]_+ | \Phi_0\rangle  
\label{eq:GAB}
\end{equation}
we obtain its Fourier transformation as
\begin{eqnarray}
G_{A,B}(\omega)&=& \langle \Phi_0 |
\hat{A}\frac{1}{\omega+E_0-\hat{H}+\rmi \eta}\hat{B}^+| \Phi_0\rangle  
\nonumber \\
&& + \langle \Phi_0 |\hat{B}^+\frac{1}{\omega-E_0+\hat{H}+\rmi \eta}\hat{A}
| \Phi_0\rangle   \; .
\label{eq:GABomega}
\end{eqnarray}
The corresponding density of states is given by
\begin{eqnarray}
D_{A,B}(\omega)&=& -\frac{1}{\pi} {\rm Im} \left[G_{A,B}(\omega)\right]   \nonumber \\
&=& \langle \Phi_0 |
\hat{A}\delta(\omega+E_0-\hat{H})\hat{B}^+ | \Phi_0\rangle   \nonumber \\
&& +  \langle \Phi_0 |
\hat{B}^+\delta(\omega-E_0+\hat{H})\hat{A}| \Phi_0\rangle   \; .
\label{eq:DABomega}
\end{eqnarray}
The first term is finite only if $\omega>0$ because $E_{\alpha}>E_0$
for the eigenenergies of $\hat{H}$. Therefore,
\begin{equation}
\int_{-\infty}^0\rmd \omega D_{A,B}(\omega)= 
\langle \Phi_0 | \hat{B}^+\hat{A}| \Phi_0\rangle 
\label{eq:expectationvalueBA}
\end{equation}
for the ground-state expectation value of the operator product 
$\hat{B}^+\hat{A}^{\vphantom{+}}$.

As an example, we give explicit expressions for the impurity occupancies,
\begin{eqnarray}
n_{b,\sigma}&=& \langle \Phi_0 | \hat{h}_{b,\sigma}^+\hat{h}_{b,\sigma}| \Phi_0\rangle
 = \int_{-\infty}^0\rmd \omega D_{b,\sigma}(\omega)  \nonumber \\
&=& -\frac{1}{\pi} \int_{-\infty}^0\rmd \omega 
{\rm Im}\left[ \frac{g_{\bar{b}}^{-1}(\omega)-H_{\bar{b},\bar{b}}(\omega)}{
\det(\omega)} \right]\; .
\label{eq:nsigmafromDOS}
\end{eqnarray}
For the hybridization matrix element we find
\begin{equation}
\langle \Phi_0 | \hat{c}_{\veck,\sigma}^+\hat{h}_{b,\sigma}| \Phi_0\rangle
 = \int_{-\infty}^0\rmd \omega D_{b,\veck}(\omega)  \; ,
\label{eq:hybfromDOS}
\end{equation}
where
\begin{equation}
D_{b,\veck}(\omega) = -\frac{1}{\pi} {\rm Im}\left[ 
G_{b,\veck}(\omega) \right]
= -\frac{1}{\pi} {\rm Im}\left[ \frac{Z_{b,\veck}(\omega)}{\det(\omega)}\right]
 \; ,
\end{equation}
see eq.~(\ref{eq:resultGBk}). 
The hybridization energy reads
\begin{eqnarray}
E_{\rm hyb}&=& \sum_{\veck,b,\sigma} V_{\veck,b}
\langle \Phi_0 | \hat{c}_{\veck,\sigma}^+\hat{h}_{b,\sigma}| \Phi_0\rangle +{\rm c.c.}
\nonumber \\
& =& -\frac{4}{\pi} \sum_b\int_{-\infty}^0\rmd \omega 
{\rm Im}
\left[ \frac{H_{b,\bar{b}}H_{\bar{b},b}(\omega)}{\det(\omega)}\right]
\nonumber \\
&& \hphantom{-\frac{1}{\pi} \sum_b}
+ {\rm Im}
\left[\frac{ H_{b,b}(\omega)(g_{\bar{b}}^{-1}(\omega)-H_{\bar{b},\bar{b}}(\omega))
}{\det(\omega)}\right]\; .\nonumber \\
\label{eq:hybenergy}
\end{eqnarray}
Note that the integrals 
in eqs.~(\ref{eq:gsenergyfromDOS}), (\ref{eq:nsigmafromDOS}) 
and~(\ref{eq:hybenergy}) 
range over a finite interval because the host bandwidth is finite.

\section{Particle-hole symmetry at half band-filling}
\label{sec:phsymm}

We are interested in the case where there is on average one electron on each 
of the impurities. This can be assured for a particle-hole symmetric 
Hamiltonian~(\ref{eq:defH}) at half band-filling. 

\subsection{Conditions}

We consider a bipartite lattice. We assume that there exists half a reciprocal
lattice vector~$\vecQ=\vecG/2$ for which
\begin{equation}
\epsilon(\veck\pm \vecQ) = -\epsilon(\veck)
\; , \; e^{\rmi \vecQ\cdot \vecR}= \left\{
\begin{array}{rcl}
1 & \hbox{if} &\vecR \in \hbox{$A$-lattice} \\
-1 & \hbox{if} & \vecR \in \hbox{$B$-lattice}
\end{array}
\right. .
\label{eq:phdemandepsk}
\end{equation}
We also assume inversion symmetry, $\epsilon(-\veck)= \epsilon(\veck)$;
recall that $\epsilon(\veck + \vecG)=\epsilon(\veck)$.
For the electron transfer matrix elements between two sites
at distance~$\vecR$ this implies
\begin{eqnarray}
t(\vecR) &=& \frac{1}{L} \sum_{\veck} e^{\rmi \veck\cdot \vecR} \epsilon(\veck)
\nonumber \\
&=& \frac{1}{L} \sum_{\veck\in {\rm MBZ}} 
\left(e^{\rmi \veck\cdot \vecR} -e^{\rmi (\vecQ-\veck) \cdot\vecR}\right)\epsilon(\veck)
\; ,
\end{eqnarray}
where MBZ is the reduced (or `magnetic') Brillouin zone that contains
only half of the vectors of the Brillouin zone.

Consequently, the electron transfer matrix elements between sites on the
same sublattice ($\vecR\in A$-lattice) ought to be imaginary, and
those between sites on different
sublattices ($\vecR\in B$-lattice) must be real,
\begin{eqnarray}
t(\vecR\in A) &=& \frac{1}{L} \sum_{\veck\in {\rm MBZ}} 
2\rmi \sin(\veck \cdot\vecR) \epsilon(\veck) \; , \nonumber \\
t(\vecR\in B) &=& \frac{1}{L} \sum_{\veck\in {\rm MBZ}} 
2\cos(\veck \cdot\vecR) \epsilon(\veck) \; .
\end{eqnarray}
However, in a solid in the absence of spin-orbit coupling, 
the tunnel amplitudes for electrons are real so 
that particle-hole symmetry actually implies $t(\vecR\in A)\equiv 0$.
In our conceptual study we do not impose this constraint.

We demand that the transfer element $t_{12}$ in
eq.~(\ref{eq:defTd}) has the same properties as $t(\vecR)$.
Therefore, we assume $t_{12}$ to be imaginary when
$\vecR_1$ and $\vecR_2$ are on the same sublattice,
and real when they are on different sublattices. This can be cast into 
the relation
\begin{equation}
t_{12}^*= -t_{12} e^{\rm i\vecQ\cdot(\vecR_1-\vecR_2)} \; .
\label{eq:phdemanddhopping}
\end{equation}
Therefore, 
\begin{equation}
\alpha_{12}^2=-e^{\rmi \vecQ(\vecR_1-\vecR_2)} 
\end{equation}
in eq.~(\ref{eq:defalpha12}).

Lastly, we demand that
\begin{equation}
V_{\veck}= V_{\vecQ-\veck}^* \; .
\label{eq:phdemandVk}
\end{equation}
Note that a purely local hybridization, $V_{\veck}\equiv V$,
must necessarily be real. 
We impose the conditions~(\ref{eq:phdemandepsk}), (\ref{eq:phdemanddhopping}),
and~(\ref{eq:phdemandVk}) to make the Hamiltonian
invariant under particle-hole transformation.

\subsection{Particle-hole transformation}

We employ 
\begin{eqnarray}
\tau_{\rm ph}\; : 
\hat{c}_{\veck,\sigma}^{\vphantom{+}} 
&\mapsto& 
\hat{\tau}_{\rm ph}^{\vphantom{+}}
\hat{c}_{\veck,\sigma}^{\vphantom{+}} 
\hat{\tau}_{\rm ph}^+=
\hat{c}_{\vecQ-\veck,\sigma}^+ 
\; , \nonumber \\
\hat{d}_{b,\sigma}^{\vphantom{+}} 
&\mapsto& 
\hat{\tau}_{\rm ph}^{\vphantom{+}}
\hat{d}_{b,\sigma}^{\vphantom{+}} 
\hat{\tau}_{\rm ph}^+=
-e^{\rm i \vecQ\cdot \vecR_b}\hat{d}_{b,\sigma}^+ 
\label{eq:phtrafodef}
\end{eqnarray}
as particle-hole transformation so that 
$\hat{n}_{b,\sigma} \mapsto 1-\hat{n}_{b,\sigma}$.
The unitary operator
$\hat{\tau}_{\rm ph}^{\vphantom{+}}$ that generates the particle-hole transformation
is provided in appendix~\ref{app:phop}.

Using eqs.~(\ref{eq:phdemanddhopping}) and~(\ref{eq:phtrafodef})
it is readily shown that
$\hat{T}_d$ and $\hat{H}_{\rm int}$ are  particle-hole symmetric,
i.e., 
$\hat{\tau}_{\rm ph}^{\vphantom{+}}\hat{T}_d\hat{\tau}_{\rm ph}^+=\hat{T}_d$ 
and 
$\hat{\tau}_{\rm ph}^{\vphantom{+}}\hat{H}_{\rm int}\hat{\tau}_{\rm ph}^+
=\hat{H}_{\rm int}$.
Moreover, eqs.~(\ref{eq:phdemandepsk}) and~(\ref{eq:phtrafodef}) ensure that 
$\hat{T}$ maps onto itself under~$\tau_{\rm ph}$.
Lastly, the hybridization $\hat{V}$ is seen to be invariant
when we use eqs.~(\ref{eq:phdemandVk}) and~(\ref{eq:phtrafodef}).
Therefore, the particle-hole transformation maps $\hat{H}$ onto itself,
\begin{equation}
\hat{\tau}_{\rm ph}^{\vphantom{+}}\hat{H}\hat{\tau}_{\rm ph}^+=\hat{H} \; .
\end{equation}
The operator for the total particle number is given by
\begin{equation}
\hat{N}=\sum_{\veck,\sigma}\hat{c}_{\veck,\sigma}^+\hat{c}_{\veck,\sigma}^{\vphantom{+}}
+ \sum_{b,\sigma} \hat{n}_{b,\sigma} \; .
\end{equation}
Under the particle-hole transformation~$\tau_{\rm ph}$
it transforms as
\begin{equation}
\tau_{\rm ph}\; : \hat{N} \mapsto 2L+4-\hat{N}  \; .
\label{eq:phtrafoN}
\end{equation}
Therefore, the particle-hole transformation~$\tau_{\rm ph}$
maps systems at and above half filling to those 
at and below half filling, and vice versa.

\subsection{Half-filled bands}

In the following we consider paramagnetic bands at half filling
where the number of electrons $N=N_{\uparrow}+N_{\downarrow}$ 
equals the (even) number of orbitals, $N=L+2$,
and $N_{\uparrow}=N_{\downarrow}=L/2+1$. Note that there are $L$ lattices
sites for the host electrons and two additional impurity orbitals
on the lattice sites $\vecR_1$ and $\vecR_2$.

At half band-filling, the non-degenerate ground state $|\Psi_0\rangle$
maps onto itself under the particle-hole transformation,
$\hat{\tau}_{\rm ph}^+|\Psi_0\rangle=|\Psi_0\rangle$.
Therefore, we find
\begin{equation}
\langle \Psi_0 | \hat{n}_{b\sigma} | \Psi_0 \rangle 
= \langle \Psi_0 | \hat{\tau}_{\rm ph}^{\vphantom{+}}\hat{n}_{b,\sigma}
\hat{\tau}_{\rm ph}^+ | \Psi_0 \rangle =
1- \langle \Psi_0 | \hat{n}_{b\sigma} | \Psi_0 \rangle \; ,
\end{equation}
i.e.,  each impurity level is exactly half filled for all
hybridizations and interaction strengths,
\begin{equation}
\langle \Psi_0 | \hat{n}_{b\sigma} | \Psi_0 \rangle =1/2 \; .
\label{eq:dsarehalffilled}
\end{equation}
Moreover, it is readily shown that the bare density of states
is symmetric,
\begin{equation}
D_{\sigma,0}(\epsilon) = \sum_{\veck} \delta(\epsilon-\epsilon(\veck))
=D_{\sigma,0}(-\epsilon) \; ,
\end{equation}
so that the Fermi energy is at $E_{\rm F}=0$ at half band-filling.

When the Hamiltonian is expressed in the $h$-basis,
we note that
\begin{equation}
\langle \Psi_0 | \hat{h}_{1,\sigma}^+\hat{h}_{1,\sigma}^{\vphantom{+}} 
+\hat{h}_{2,\sigma}^+\hat{h}_{2,\sigma}^{\vphantom{+}} |\Psi_0\rangle=1
\label{eq:btwoandbonedensities}
\end{equation}
at half band-filling. 
Moreover, the particle-hole transformation implies
\begin{equation}
\tau_{\rm ph}\, : 
\hat{h}_{b,\sigma}^{\vphantom{+}} \mapsto e^{-\rmi \gamma_{12}}
\hat{h}_{\bar{b},\sigma}^+ 
\; , \; e^{-\rmi \gamma_{12}}=-\alpha_{12} e^{\rmi \vecQ\cdot \vecR_1} \; ,
\end{equation}
where we used the notation $\bar{1}=2$, $\bar{2}=1$.
Therefore, par\-ticle-hole symmetry at half band-filling leads to
\begin{equation}
\langle \Psi_0 | 
\hat{h}_{1,\sigma}^+\hat{h}_{2,\sigma}^{\vphantom{+}} 
|\Psi_0\rangle
=
\langle \Psi_0 | 
\hat{h}_{2,\sigma}^{\vphantom{+}}\hat{h}_{1,\sigma}^+
|\Psi_0\rangle \;, 
\end{equation}
so that there is no hybridization between the $h$-orbitals at half band-filling,
\begin{equation}
\langle \Psi_0 | 
\hat{h}_{1,\sigma}^+\hat{h}_{2,\sigma}^{\vphantom{+}} 
|\Psi_0\rangle
=0 \; .
\label{eq:h12andh21are zero}
\end{equation}

Note that, for $\vecR_1=\vecR_2$, we have $\alpha_{11}^2=-1$ so that 
$V_{\veck,1}=V_{\veck,2}\neq 0$ in eq.~(\ref{eq:hybridizationsforh}).
Even in this case, both impurity levels couple to the bath electrons and 
the separation between `odd' and `even' 
channels does not lead to an inert impurity level as in~\cite{ReinhardSatoshi}.

\section{Tight-binding model with local hybridization}
\label{sec:TBmodel}

To arrive at definite results, we consider spin-1/2 host electrons
that move between nearest neighbors on 
a three-dimensional simple-cubic lattice. Moreover, we assume that the
hybridization between host levels and impurity levels and the
direct transfer between the impurities are weak (RKKY limit).
In this limit we provide explicit expressions for the ground-state energy, 
impurity density, and impurity contribution to the spectral function.

\subsection{Model setup}

First, we specify the dispersion relation of the host electrons and the hybridization
matrix elements and calculate 
the host-electron density-density correlation function.
We work out the host Green functions
and hybridization functions under the assumption that
the hybridization is weak and purely local.

\subsubsection{Host electron dispersion relation and hybridization} 

The dispersion relation is given by
\begin{equation}
\epsilon(\veck)= -\frac{W}{6} \bigl(\cos(k_x)+\cos(k_y)+\cos(k_z)\bigr)\; ,
\label{eq:dispersion}
\end{equation}
where we set the lattice spacing to unity, $a\equiv 1$.
Henceforth we use the bandwidth as our energy unit, $W\equiv 1$.
The nesting vector is $\vecQ=(\pi,\pi,\pi)$, $\epsilon(\vecQ-\veck)=-\epsilon(\veck)$.

We place the first impurity onto
the origin of the $A$-lattice, $\vecR_1=\veczero$, 
and write $\vecR_2\equiv \vecR$ for simplicity. 
Then, we have ($\alpha_{12}\equiv \alpha_{\vecR}$)
\begin{equation}
\alpha_{\vecR}^2=-\exp(-\rmi \vecQ\cdot \vecR)=
(-1)^{(R_x+R_y+R_z)+1}\; .
\end{equation}
We use a purely local hybridization $V_{\veck}=V$ so that
\begin{equation}
V_{\veck,1}
= \frac{V}{\sqrt{2}}(1-\alpha_{\vecR} e^{-\rmi \veck\cdot \vecR})
\,,\,
V_{\veck,2}= \frac{V}{\sqrt{2}}(\alpha_{\vecR}^*+ e^{-\rmi \veck\cdot \vecR}) \; .
\label{eq:deflocalV}
\end{equation}

\subsubsection{Host electron density-density correlation function}

Since the RKKY interaction between the impurities is mediated by
the host electrons it is instructive to study the density-density correlation function
in the host electrons' Fermi sea.
The correlation function between like spins and $\vecR\neq \veczero$ is given by
($\hat{n}_{c,\vecR,\sigma}=
\hat{c}_{\vecR,\sigma}^+\hat{c}_{\vecR,\sigma}^{\vphantom{+}}$)
\begin{eqnarray}
C_{\uparrow}^{\rm nn}(\vecR)&=&
\frac{1}{L}\sum_{\vecR'}\langle {\rm FS}| \hat{n}_{c,\vecR',\uparrow}
 \hat{n}_{c,\vecR'+\vecR,\uparrow}| {\rm FS} \rangle -n_{\uparrow}^2\nonumber \\
&=& - \left| 
\frac{1}{L} \sum_{\veck}e^{\rmi \veck\cdot\vecR} n_{\veck,\uparrow}^{\rm FS}
\right|^2 \; ,
\end{eqnarray}
where $n_{\uparrow}=1/2$ and $n_{\veck,\uparrow}^{\rm FS}=1-\Theta[\epsilon(k)]$
is unity in the reduced Brillouin zone.
For $\vecR\in \hbox{$A$-lattice}$ we have $\exp(\rmi \vecQ \cdot\vecR)=1$
so that
\begin{eqnarray}
\frac{1}{L} \sum_{\veck}e^{\rmi \veck\cdot\vecR} n_{\veck,\uparrow}^{\rm FS}
&=& \int_{-\infty}^0\rmd \epsilon
\frac{1}{L}\sum_{\veck} \delta(\epsilon-\epsilon(\veck)) 
e^{\rmi \veck\cdot\vecR} \nonumber\\
&=&\int_{-\infty}^0\rmd \epsilon\frac{1}{2L}\sum_{\veck} 
\left[\delta(\epsilon-\epsilon(\veck)) e^{\rmi \veck\cdot\vecR} 
\right. \nonumber \\
&&\hphantom{\int_{-\infty}^0\rmd \frac{1}{2L}}
\left.+\delta(\epsilon+\epsilon(\veck)) 
e^{\rmi (\vecQ-\veck)\cdot\vecR} \right]
\nonumber\\
&=& \int_{-\infty}^{\infty}\rmd \epsilon
\frac{1}{2L}\sum_{\veck}\delta(\epsilon-\epsilon(\veck)) 
e^{\rmi \veck\cdot\vecR} \nonumber \\
&=& \frac{1}{2}\delta_{\vecR,\veczero}=0
\end{eqnarray}
because $\vecR\neq \veczero$.
Since the host electrons mediate the RKKY interaction but do not show any
non-trivial correlations for $\vecR\in \hbox{$A$-lattice}$, the RKKY coupling 
does not lead to a level splitting, as shown in section~\ref{sec:RinAimpurityDOS}.

\subsubsection{Host-electron Green functions}
First, we define two distance-dependent density of states
$D_{A,B}(\omega;\vecR)$
and their Hilbert-transformed $\Lambda_{A,B}(\omega;\vecR)$
that naturally appear in the hybridization functions, 
see section~\ref{sec:hybfunctionsexplicit}.
To this end, we first consider 
\begin{eqnarray}
D_A(\omega;\vecR)&=&
\sum_{\veck}\frac{\delta(\omega-\epsilon(\veck))}{2L}
(e^{\rmi \veck\cdot \vecR}+e^{-\rmi \vecQ\cdot\vecR} e^{-\rmi \veck\cdot \vecR})\; ,
\nonumber \\
\Lambda_A(\omega;\vecR)&=& \dashint_{-1/2}^{1/2} \rmd \omega'
\frac{D_A(\omega';\vecR)}{\omega-\omega'} \; .
\label{eq:defDALamA}
\end{eqnarray}
These quantities are the real and imaginary parts of 
the complex functions
\begin{eqnarray}
H_A(\omega;\vecR)&=&\Lambda_A(\omega;\vecR)-\rmi \pi D_A(\omega;\vecR)\nonumber\\
&=& \frac{1}{2L}\sum_{\veck} 
\frac{(e^{\rmi \veck\cdot \vecR}+e^{-\rmi \vecQ\cdot\vecR} e^{-\rmi \veck\cdot \vecR})}{
\omega-\epsilon(\veck)+\rmi \eta} \; .
\label{eq:defAlatticelGF} 
\end{eqnarray}
This relation shows that $H_{A;\vecR}(\omega;\vecR)$ is finite only if 
$\vecR\in\hbox{$A$-lattice}$. For this reason, we added the redundant index~$A$.
For our dispersion~(\ref{eq:dispersion}) we have from the Fourier transformation
\begin{eqnarray}
H_A(\omega;\vecR)&=&(-\rmi)\int_0^{\infty}\rmd t e^{\rmi (\omega+\rmi \eta) t}
\nonumber \\
&& \times
\frac{1}{2L}\sum_{\veck} e^{-\rmi \epsilon(\veck)t}
\left(e^{\rmi \veck\cdot\vecR}+e^{-\rmi \vecQ\cdot \vecR}e^{-\rmi \veck\cdot\vecR}\right)
\nonumber \\
&=& \delta_{\vecR\in A}
(-1)^{(R_x+R_y+R_z)/2} (-\rmi)\nonumber \\
&& \times \!\int_0^{\infty}\rmd t e^{\rmi \omega t}
J_{R_x}\Bigl(\frac{t}{6}\Bigr)J_{R_y}\Bigl(\frac{t}{6}\Bigr)J_{R_z}\Bigl(\frac{t}{6}\Bigr)\;,
\nonumber \\
\label{eq:defHA}
\end{eqnarray}
where the first factor implies that $\vec{R}\in \hbox{$A$-lattice}$, i.e.,
$R_x+R_y+R_z$ must be an even integer.
Here, $J_n(x)=(-1)^nJ_n(-x)$ is the Bessel function of integer order~$n$.
Consequently, we have from the real and imaginary parts
\begin{eqnarray}
\Lambda_A(\omega;\vecR)&=& 
\delta_{\vecR\in A}
(-1)^{(R_x+R_y+R_z)/2} \nonumber \\
&& \times \!\!\int_0^{\infty} \!\rmd t \sin(\omega t)
J_{R_x}\Bigl(\frac{t}{6}\Bigr)J_{R_y}\Bigl(\frac{t}{6}\Bigr)J_{R_z}\Bigl(\frac{t}{6}\Bigr),
\nonumber \\
D_A(\omega;\vecR)&=&
\delta_{\vecR\in A}
(-1)^{(R_x+R_y+R_z)/2} \nonumber \\
&& \times \!\!\int_0^{\infty}\!\frac{\rmd t}{\pi} \cos(\omega t)
J_{R_x}\Bigl(\frac{t}{6}\Bigr)J_{R_y}\Bigl(\frac{t}{6}\Bigr)J_{R_z}\Bigl(\frac{t}{6}\Bigr)
\nonumber \\
\label{eq:resultDALamA}
\end{eqnarray}
for $\vecR\in\hbox{$A$-lattice}$.
For later use we denote
$D_{\veczero}(\omega)=D_A(\omega;\veczero)$ and
$\Lambda_{\veczero}(\omega)=\Lambda_A(\omega;\veczero)$.

Next, we define 
\begin{eqnarray}
D_B(\omega;\vecR)&=&
\sum_{\veck}\frac{\delta(\omega-\epsilon(\veck))}{2L}
(\alpha_{\vecR} e^{-\rmi \veck\cdot \vecR}+\alpha_{\vecR}^* e^{\rmi \veck\cdot \vecR})\; ,
\nonumber \\
\Lambda_B(\omega;\vecR)&=& \dashint_{-1/2}^{1/2} \rmd \omega'
\frac{D_B(\omega';\vecR)}{\omega-\omega'} \; .
\label{eq:defDBLamB}
\end{eqnarray}
These quantities are the real and imaginary parts of 
the complex functions
\begin{eqnarray}
H_B(\omega;\vecR)&=&\Lambda_B(\omega;\vecR)-\rmi \pi D_B(\omega;\vecR)
\nonumber \\
&=& \frac{1}{2L}\sum_{\veck} 
\frac{\alpha_{\vecR} e^{-\rmi \veck\cdot\vecR}+\alpha_{\vecR}^* e^{\rmi \veck\cdot\vecR}}{
\omega-\epsilon(\veck)+\rmi \eta} \; .
\label{eq:defBlatticelGF} 
\end{eqnarray}
Since $\alpha_{\vecR}^*=-\alpha_{\vecR}$ for
$\vecR\in\hbox{$A$-lattice}$, 
$H_{B;\vecR}(\omega;\vecR)$ is finite only if $\vecR\in\hbox{$B$-lattice}$.
Again, we indicate this by the redundant index~$B$.

For our dispersion~(\ref{eq:dispersion}) we have from the Fourier transformation
\begin{eqnarray}
H_B(\omega;\vecR)&=&(-\rmi)\int_0^{\infty}\rmd t e^{\rmi (\omega+\rmi \eta) t}
\nonumber \\
&& \times
\frac{1}{2L}\sum_{\veck} e^{-\rmi \epsilon(\veck)t}
\left(\alpha_{\vecR} e^{-\rmi \veck\cdot\vecR}+\alpha_{\vecR}^* e^{\rmi \veck\cdot\vecR}\right)
\nonumber \\
&=& \delta_{\vecR\in B}
(-\rmi){\rm i}^{R_x+R_y+R_z}\nonumber\\
&& \times\int_0^{\infty}\rmd t e^{\rmi \omega t}
J_{R_x}\Bigl(\frac{t}{6}\Bigr)J_{R_y}\Bigl(\frac{t}{6}\Bigr)J_{R_z}\Bigl(\frac{t}{6}\Bigr)
\; ,\nonumber \\
\label{eq:defHB}
\end{eqnarray}
where the first factor implies that $\vec{R}\in \hbox{$B$-lattice}$, i.e.,
$R_x+R_y+R_z$ must be an odd integer.
Consequently, we have from the real and imaginary parts
\begin{eqnarray}
\Lambda_B(\omega;\vecR)&=& 
\delta_{\vecR\in B}
(-1)^{(R_x+R_y+R_z+3)/2}\nonumber \\
&& \times  \!\!\int_{0}^{\infty}\!\!\rmd t \cos(\omega t)
J_{R_x}\Bigl(\frac{t}{6}\Bigr)J_{R_y}\Bigl(\frac{t}{6}\Bigr)J_{R_z}\Bigl(\frac{t}{6}\Bigr),
\nonumber \\
D_B(\omega;\vecR)&=&
\delta_{\vecR\in B}
(-1)^{(R_x+R_y+R_z+1)/2}\nonumber\\
&& \times\!\!\int_{0}^{\infty}\!\frac{\rmd t}{\pi}\sin(\omega t)
J_{R_x}\Bigl(\frac{t}{6}\Bigr)J_{R_y}\Bigl(\frac{t}{6}\Bigr)J_{R_z}\Bigl(\frac{t}{6}\Bigr)
\nonumber \\
\label{eq:resultDBLamB}
\end{eqnarray}
for $\vecR\in\hbox{$B$-lattice}$.

\subsubsection{Hybridization functions}
\label{sec:hybfunctionsexplicit}
We have 
\begin{eqnarray}
V_{\veck,1} V_{\veck,1}^* &=& 
\frac{V^2}{2}(1-\alpha_{\vecR} e^{-\rmi \veck\cdot \vecR})
(1-\alpha_{\vecR}^* e^{\rmi \veck\cdot \vecR})\nonumber\\
&=& 
V^2(1-\frac{\alpha_{\vecR}}{2} e^{-\rmi \veck\cdot \vecR}
-\frac{\alpha_{\vecR}^*}{2} e^{\rmi \veck\cdot \vecR})
\; .
\end{eqnarray}
Thus, we obtain from eqs.~(\ref{eq:defHbbprime}) and~(\ref{eq:deflocalV})
\begin{eqnarray}
H_{1,1}(\omega;\vecR)&=& \frac{V^2}{L}\sum_{\veck}
\frac{1}{\omega-\epsilon(\veck)+\rmi\eta} \nonumber \\
&& - \frac{V^2}{2L}\sum_{\veck}
\frac{\alpha_{\vecR} e^{-\rmi \veck\cdot \vecR}+\alpha_{\vecR}^* e^{\rmi \veck\cdot \vecR}}{
\omega-\epsilon(\veck)+\rmi\eta} \nonumber \\
&=& V^2\bigl(\Lambda_{\veczero}(\omega)- \Lambda_B(\omega;\vecR)\bigr)
\nonumber \\
&& -\rmi \pi V^2 \bigl(D_{\veczero}(\omega)- D_B(\omega;\vecR)\bigr) \; .
\end{eqnarray}
Next, 
\begin{eqnarray}
V_{\veck,2} V_{\veck,2}^* &=& 
\frac{V^2}{2L}(\alpha_{\vecR}^*+ e^{-\rmi \veck\cdot \vecR})(\alpha_{\vecR}
+ e^{\rmi \veck\cdot \vecR})\nonumber \\
&=& 
\frac{V^2}{L}(1+\frac{\alpha_{\vecR}}{2} e^{-\rmi \veck\cdot \vecR}
+\frac{\alpha_{\vecR}^*}{2} e^{\rmi \veck\cdot \vecR})
\; .
\end{eqnarray}
Thus, we obtain 
\begin{eqnarray}
H_{2,2}(\omega;\vecR)&=& 
V^2\bigl(\Lambda_{\veczero}(\omega)+ \Lambda_B(\omega;\vecR)\bigr)\nonumber \\
&& -\rmi \pi V^2 \bigl(D_{\veczero}(\omega)+ D_B(\omega;\vecR)\bigr) \; .
\end{eqnarray}
Finally,
\begin{eqnarray}
V_{\veck,1} V_{\veck,2}^* &=& 
\frac{V^2}{2}(1-\alpha_{\vecR} e^{-\rmi \veck\cdot \vecR})(\alpha_{\vecR}
+ e^{\rmi \veck\cdot \vecR})\nonumber \\
&=& 
\frac{V^2}{2}(e^{\rmi \veck\cdot \vecR}
+e^{-\rmi \vecQ\cdot \vecR} e^{-\rmi \veck\cdot \vecR})
\; .
\end{eqnarray}
Therefore,
\begin{equation}
H_{1,2}(\omega;\vecR)= V^2\Lambda_A(\omega;\vecR)
-\rmi \pi V^2 D_A(\omega;\vecR)\; ,
\end{equation}
and since
\begin{eqnarray}
V_{\veck,2} V_{\veck,1}^* &=& 
\frac{V^2}{2}(1-\alpha_{\vecR}^* e^{\rmi \veck\cdot \vecR})
(\alpha_{\vecR}^*+ e^{-\rmi \veck\cdot \vecR})\nonumber \\
&=& 
\frac{V^2}{2}e^{\rmi \vecQ\cdot \vecR} (e^{\rmi \veck\cdot \vecR}
+e^{-\rmi \vecQ\cdot \vecR} e^{-\rmi \veck\cdot \vecR}) \;, 
\end{eqnarray}
we find
\begin{equation}
H_{2,1}(\omega;\vecR)= H_{1,2}(\omega;\vecR) \; .
\end{equation}

\subsubsection{RKKY limit of small hybridization}
For $V\ll1$ and $|t_{12}|\ll 1$, 
the relevant frequency range for the occupation densities and
the hybridization will turn out to be also small, $|\omega|\ll 1$.
Therefore, we keep only the leading-order terms
in $D_{A,B}(\omega;\vecR)$ around $\omega=0$.

For $\omega\to 0$ we have 
\begin{eqnarray}
D_{A;\vecR}(\omega\to 0)&=& s_{\vecR}d_{\vecR}\; ,\nonumber  \\
d_{\vecR}&=&
\int_{0}^{\infty}\frac{\rmd t}{\pi}J_{R_x}\Bigl(\frac{t}{6}\Bigr)
J_{R_y}\Bigl(\frac{t}{6}\Bigr)J_{R_z}\Bigl(\frac{t}{6}\Bigr)\, ,\nonumber\\
s_{\vecR}&=&(-1)^{(R_x+R_y+R_z)/2}\; ,
\end{eqnarray}
and 
\begin{eqnarray}
\Lambda_A(\omega \to 0;\vecR)&=& \omega s_{\vecR} \ell_{\vecR}\; , 
\nonumber \\
\ell_{\vecR} &=&
\int_{0}^{\infty}\rmd t\, t
J_{R_x}\Bigl(\frac{t}{6}\Bigr)J_{R_y}\Bigl(\frac{t}{6}\Bigr)J_{R_z}\Bigl(\frac{t}{6}\Bigr)
\, ,\nonumber\\
\end{eqnarray}
where it is implicitly understood that $R_x+R_y+R_z$ must be an even integer.
The integrals for $d_{\vecR}$ and $\ell_{\vecR}$
can be calculated numerically with {\sc Mathematica}~\cite{Mathematica}.

Moreover, for $\omega\to 0$ we have 
\begin{eqnarray}
D_B(\omega\to 0;\vecR)&=& \tilde{s}_{\vecR}
\frac{\omega }{\pi}\ell_{\vecR}
\;, \\
\Lambda_{B;\vecR}(\omega\to 0)&=& -\pi \tilde{s}_{\vecR}d_{\vecR} \; ,
\nonumber \\
\tilde{s}_{\vecR}&=&(-1)^{(R_x+R_y+R_z+1)/2} \; ,
\end{eqnarray}
where it is implicitly understood that $R_x+R_y+R_z$ must be an odd integer.

\subsection{Particle density}

We recall that the first impurity is at the origin that belongs to
sublattice~$A$ by definition, and the second impurity lies at~$\vecR$.
We treat the two cases,  $\vecR \in \hbox{$A$/$B$-lattice}$
separately; it is implicitly understood in the following 
that $\vecR$ lies in the corresponding sublattice. 

\subsubsection{$\vecR \in \hbox{$B$-lattice}$}

We start with the simpler of the two cases.
We use 
$D_{\veczero}(\omega)\approx d_{\veczero}$,
$\Lambda_{\veczero}(\omega)\approx \omega\ell_{\veczero}$,
and
$D_B(\omega;\vecR)\approx (\omega/\pi)\tilde{s}_{\vecR}\ell_{\vecR}$,
$\Lambda_B(\omega;\vecR)\approx-\pi\tilde{s}_{\vecR}d_{\vecR}$.
Thus, we obtain
\begin{eqnarray}
H_{1,1}(\omega) &\approx& V^2(\omega\ell_{\veczero}+\pi\tilde{s}_{\vecR}d_{\vecR})
-\rmi V^2(\pi d_{\veczero}-\omega\tilde{s}_{\vecR}\ell_{\vecR})\nonumber \; , \\
H_{2,2}(\omega) &\approx& V^2(\omega\ell_{\veczero}-\pi\tilde{s}_{\vecR}d_{\vecR})
-\rmi V^2(\pi d_{\veczero}+\omega\tilde{s}_{\vecR}\ell_{\vecR})\nonumber \; , \\
H_{1,2}(\omega) &=& 0\nonumber \; , \\
H_{2,1}(\omega) &=& 0\; .
\end{eqnarray}
Since the mixing terms vanish, we can considerably simplify $n_{1,\uparrow}$ to
\begin{eqnarray}
n_{1,\uparrow}&=& -\frac{1}{\pi} \int_{-1/2}^0 \rmd \omega {\rm Im}
\left[ \frac{1}{\omega+|t_{12}|-H_{1,1}(\omega)}
\right] \nonumber \\
&\approx& 
\frac{1}{\pi} \int_{-C}^0 \rmd \omega 
 \frac{V^2(\pi d_{\veczero}-\omega\tilde{s}_{\vecR}\ell_{\vecR})}{P_B(\omega)}
\label{eq:nsigmaintegral}
\end{eqnarray}
with
\begin{eqnarray}
P_B(\omega)&=&
[\omega+|t_{12}|-V^2(\omega\ell_{\veczero}+\pi\tilde{s}_{\vecR}d_{\vecR})]^2
\nonumber \\
&& +[V^2(\pi d_{\veczero}-\omega\tilde{s}_{\vecR}\ell_{\vecR})]^2 \; ,
\end{eqnarray}
where $C>0$ is of order unity. Corrections are small, of order $V^2$.
For $|t_{12}|,V\to 0$ it is useful to set
\begin{equation}
\omega= V^2 x\quad , \quad |t_{12}|=V^2 \bar{t} \quad, \quad C/V^2\to \infty \; .
\label{eq:defxtbar}
\end{equation}
Ignoring terms of order $V^2$ and higher, we arrive at
\begin{equation}
n_{1,\uparrow}\approx d_{\veczero}  \int_{-\infty}^0 \rmd x [\tilde{P}_B(x)]^{-1}
\end{equation}
with
\begin{equation}
\tilde{P}_B(x)=
x^2+2x(\bar{t}-\pi\tilde{s}_{\vecR}d_{\vecR})
+(\bar{t}-\pi\tilde{s}_{\vecR}d_{\vecR})^2
+(\pi d_{\veczero})^2 \; .
\end{equation}
The integral is elementary and gives
\begin{equation}
n_{1,\uparrow}\approx \frac{1}{\pi} \left[ 
\arctan\left(\frac{\bar{t}-\pi\tilde{s}_{\vecR}d_{\vecR}}{\pi d_{\veczero}}\right) 
+\frac{\pi}{2}\right] \; .
\label{eq:noftbarrelationB}
\end{equation}
For an impurity coupling~$|t_{12}|$ of order unity, the large value for 
$\bar{t}=|t_{12}|/V^2$
leads to $n_{1,\uparrow}\approx 1$, irrespective of~$\vecR$.
For this reason, it is important to have a small direct coupling
of the impurities,  $|t_{12}|={\cal O}(V^2)$, to 
obtain a RKKY behavior, see sections~\ref{sec:levelRKKYsplitting}
and~\ref{sec:RKKYenegry}.

\subsubsection{$\vecR \in \hbox{$A$-lattice}$}
\label{subsec:RinAapprox}

We use 
$D_{\veczero}(\omega)\approx d_{\veczero}$,
$\Lambda_{\veczero}(\omega)\approx \omega\ell_{\veczero}$,
and
$D_A(\omega;\vecR)\approx s_{\vecR}d_{\vecR}$,
$\Lambda_A(\omega;\vecR)\approx\omega s_{\vecR}\ell_{\vecR}$
to find
\begin{eqnarray}
H_{1,1}(\omega) &\approx& V^2\omega\ell_{\veczero}
-\rmi \pi V^2d_{\veczero}\nonumber \; , \\
H_{2,2}(\omega) &= & H_{1,1}(\omega) \nonumber \; , \\
H_{1,2}(\omega) &\approx& V^2\omega s_{\vecR} \ell_{\vecR}
-\rmi \pi V^2 s_{\vecR} d_{\vecR}
\nonumber \; , \\
H_{2,1}(\omega) &=& H_{1,2}(\omega)\; .
\end{eqnarray}
Then, 
\begin{eqnarray}
\det(\omega)&\approx& (\omega-V^2\omega\ell_{\veczero}
+\rmi \pi V^2d_{\veczero})^2-|t_{12}|^2
\nonumber \\ 
&& -[H_{1,2}(\omega)]^2\; , \nonumber\\
{\rm Re}[\det(\omega)]
&=& \omega^2(1-V^2\ell_{\veczero})^2-(\pi V^2d_{\veczero})^2 
-|t_{12}|^2 \nonumber\\
&& +(\pi V^2 d_{\vecR})^2-(V^2\omega \ell_{\vecR})^2\; , \\
{\rm Im}[\det(\omega)]&=& 
2\pi \omega V^2d_{\veczero}(1-V^2\ell_{\veczero})
+2\pi \omega V^4 \ell_{\vecR} d_{\vecR}\; .\nonumber
\end{eqnarray}
The density simplifies to
\begin{eqnarray}
n_{1,\uparrow}&\approx& -\frac{1}{\pi} \int_{-C}^0 \rmd \omega
\left[ \frac{\pi V^2d_{\veczero}{\rm Re}[\det(\omega)]}{P_A(\omega)}\right.
\nonumber\\
&& \hphantom{ -\frac{1}{\pi} }
\left. -\frac{(\omega(1-V^2\ell_{\veczero})-|t_{12}|){
\rm Im}[\det(\omega)]}{J_A(\omega)}
\right]
\label{eq:calculatenA}
\end{eqnarray}
with
\begin{equation}
J_A(\omega)=
({\rm Re}[\det(\omega)])^2+({\rm Im}[\det(\omega)])^2
\end{equation}
and a high-energy cut-off, $C={\cal O}(1)$.
For $|t_{12}|,V\to 0$ we use eq.~(\ref{eq:defxtbar})
and neglect higher-order terms like $V^2\omega\ll V^2$. Then,
\begin{eqnarray}
{\rm Re}[\det(x)]&=& V^4\left[
x^2-\bar{t}^2 +(\pi d_{\vecR})^2-(\pi d_{\veczero})^2\right] \; ,
\nonumber \\
{\rm Im}[\det(x)]&=& 2\pi x V^4d_{\veczero} \; .
\end{eqnarray}
The particle density becomes formally independent of $V^2$ and reads
\begin{equation}
n_{1,\uparrow}\approx d_{\veczero} \int_{-\infty}^0 \rmd x
\frac{x^2+\bar{t}^2(1 +p_{\vecR})-2 x\bar{t}}{
(x^2-\bar{t}^2(1+p_{\vecR}))^2+(2\pi x d_{\veczero})^2}
\end{equation}
with 
\begin{equation}
p_{\vecR}=\frac{\pi^2(d_{\veczero}^2-d_{\vecR}^2)}{\bar{t}^2}>0\; .
\label{eq:defalpha}
\end{equation}
We split the integral into even and odd terms,
and substitute $y=x/\bar{t}$ in the integral over even powers and 
$\lambda=x^2/\bar{t}^2$ in the integral over odd powers,
\begin{eqnarray}
n_{1,\uparrow}
&=& 
\frac{d_{\veczero}}{2\bar{t}}
\int_{-\infty}^{\infty}
\frac{(y^2+(1+p_{\vecR}))\rmd y}{y^4-2y^2(1+p_{\vecR})+
(1+p_{\vecR})^2+y^2\kappa^2} \nonumber \\
&& +  \frac{d_{\veczero}}{\bar{t}}\int_0^{\infty}
\frac{ \rmd \lambda}{\lambda^2+\lambda(\kappa^2-2(1+p_{\vecR}))+(1+p_{\vecR})^2}
\nonumber\\
\end{eqnarray}
with $\kappa= 2\pi d_{\veczero}/\bar{t}$.
The first term integrates to 1/2. The second term is elementary.
We define
\begin{equation}
\tilde{\kappa}_{\vecR}= 
\frac{\kappa}{\sqrt{1+p_{\vecR}}}
=\frac{2\pi d_{\veczero}}{\sqrt{\bar{t}^2+\pi^2(d_{\veczero}^2-d_{\vecR}^2)\vphantom{A^A}}}
\label{eq:defkappatilde}
\end{equation}
and find with the substitution $\lambda=(1+p_{\vecR})\mu$
\begin{eqnarray}
n_{1,\uparrow}&=& \frac{1}{2}
+ \frac{\kappa}{2\pi(1+p_{\vecR})} \int_0^{\infty} \rmd \mu
\frac{1}{\mu^2+\mu(\tilde{\kappa}_{\vecR}^2-2)+1} \nonumber\\
&=& \frac{1}{2} +\frac{1}{2\pi \sqrt{1+p_{\vecR}}} G(\tilde{\kappa}_{\vecR})\; ,
\label{eq:tofnAlattice}
\end{eqnarray}
where $G(0<x<2)\equiv G_1(x)$ and
$G(x\geq 2)\equiv G_2(x)$ with
\begin{eqnarray}
G_1(x)&=&\frac{\pi}{\sqrt{4-x^2}} - \frac{2}{\sqrt{4-x^2}} 
\arctan\left(\frac{x^2-2}{x\sqrt{4-x^2}}\right) \;, \nonumber \\
G_2(x)&=& 
-\frac{1}{\sqrt{x^2-4}}\ln \left( 
\frac{x^2-2-x\sqrt{x^2-4}}{
x^2-2+x\sqrt{x^2-4}}
\right) \; .
\end{eqnarray}

\subsection{Impurity contribution to the density of states}

In the RKKY limit, $|t_{12}|, V\ll 1$, we can find an explicit expression for
the impurity-contribution to the density of states.
Again, we treat the two cases,  $\vecR \in \hbox{$A$/$B$-lattice}$
separately.

\subsubsection{$\vecR \in \hbox{$B$-lattice}$}

Now that we have $H_{1,2}=H_{2,1}=0$,
$H_{1,1}(\omega)=V^2\pi\tilde{s}_{\vecR}d_{\vecR} -\rmi V^2\pi d_{\veczero}$,
$H_{2,2}(\omega)=-V^2\pi\tilde{s}_{\vecR}d_{\vecR} -\rmi V^2\pi d_{\veczero}$,
$H_{1,1}'(\omega)=V^2\ell_{\veczero}+\rmi V^2\tilde{s}_{\vecR}\ell_{\vecR}$,
and
$H_{2,2}'(\omega)=V^2\ell_{\veczero}-\rmi V^2\tilde{s}_{\vecR}\ell_{\vecR}$,
we find
\begin{eqnarray}
V^2 D_{1,\sigma}(x) &=& 
\frac{d_{\veczero}}{(x+\bar{t}-\pi\tilde{s}_{\vecR}d_{\vecR})^2+(\pi d_{\veczero})^2}
\; , \nonumber \\
V^2D_{2,\sigma}(x) &=& 
\frac{d_{\veczero}}{(x-\bar{t}+\pi\tilde{s}_{\vecR}d_{\vecR})^2+(\pi d_{\veczero})^2}
\label{eq:twoLorentzians}
\end{eqnarray}
after scaling with $V^2$. The contribution of the two impurities
is given by two Lorentz-peaks at $x=\pm(\bar{t}-\pi\tilde{s}_{\vecR}d_{\vecR})$
with $\pi d_{\veczero}$ as half-width at half maximum.

Due to the hybridization with the host electrons, we find the contribution
\begin{eqnarray}
\Delta D_{\sigma}^{\rm host}(x) &=& 
{\rm Im}\left[
\frac{V^2(\ell_{\veczero}+\rmi\tilde{s}_{\vecR}\ell_{\vecR})/\pi}{\omega+
|t_{12}|-V^2\pi\tilde{s}_{\vecR}d_{\vecR}
+\rmi \pi V^2 d_{\veczero}}\right]\nonumber \\
&& +
{\rm Im}
\left[
\frac{V^2(\ell_{\veczero}-\rmi \tilde{s}_{\vecR}\ell_{\vecR})/\pi}{\omega
-|t_{12}| +V^2\pi\tilde{s}_{\vecR}d_{\vecR}+\rmi \pi V^2 d_{\veczero}}
\right] \nonumber \\
&=& 
\frac{\tilde{s}_{\vecR}\ell_{\vecR}(x+\bar{t}-\pi\tilde{s}_{\vecR}d_{\vecR})/\pi
-d_{\veczero}\ell_{\veczero}}{
(x+\bar{t}-\pi\tilde{s}_{\vecR}d_{\vecR})^2+(\pi d_{\veczero})^2}\nonumber \\
&& -
\frac{\tilde{s}_{\vecR}\ell_{\vecR}(x-\bar{t}+\pi\tilde{s}_{\vecR}d_{\vecR})/\pi
+d_{\veczero}\ell_{\veczero}}{
(x-\bar{t}+\pi\tilde{s}_{\vecR}d_{\vecR})^2+(\pi d_{\veczero})^2}\,. \nonumber \\
\end{eqnarray}
Apparently, the host contribution is of order $V^2$ smaller than 
the two Lorentzian peaks in eq.~(\ref{eq:twoLorentzians})
and can be ignored for small~$V$.
Note that $\Delta D_{\sigma}^{\rm host}(|x|\gg1)\sim 1/x^2$ because
the terms proportional to $(x/\pi)\tilde{s}_{\vecR}\ell_{\vecR}$ cancel each other
for large $|x|$.

\subsubsection{$\vecR \in \hbox{$A$-lattice}$}
\label{sec:RinAimpurityDOS}

Likewise, $\Delta D_{\sigma}^{\rm host}(\omega)$
is much smaller than the contributions
$D_b(\omega)$ to the impurity density of states.
We find as dominant terms
\begin{eqnarray}
V^2 D_{1,\sigma}(x) &=& d_{\veczero}
\frac{x^2-2x\bar{t}+\bar{t}^2(1+p_{\vecR})}{
[x^2-\bar{t}^2(1+p_{\vecR})]^2+(2\pi xd_{\veczero})^2}
\; , \nonumber \\[3pt]
V^2D_{2,\sigma}(x) &=& d_{\veczero}
\frac{x^2+2x\bar{t}+\bar{t}^2(1+p_{\vecR})}{
[x^2-\bar{t}^2(1+p_{\vecR})]^2+(2\pi xd_{\veczero})^2}
\label{eq:twopeaksA}
\end{eqnarray}
after scaling with $V^2$ and with $p_{\vecR}$ from eq.~(\ref{eq:defalpha}).

\subsection{Ground-state energy}

Accordingly, the ground-state energy simplifies in the RKKY limit, and
we find explicit expressions to order ${\cal O}(V^2\ln(V^2), V^2)$.

\subsubsection{$\vecR \in \hbox{$B$-lattice}$}

Since the contribution from $\Delta D_{\sigma}^{\rm host}(\omega)$
are a factor $V^2$ smaller than those from $D_b(\omega)$,
we have to order $V^2\ln(V^2)$ and $V^2$
\begin{eqnarray}
\Delta E_0 &=& 2|t_{12}|+\Delta\tilde{E}_0\; , \nonumber \\
\Delta \tilde{E}_0 &\approx &
2 \int_{-\infty}^0 \rmd \omega \sum_b \omega D_b(\omega)  \; .
\end{eqnarray}
With the expressions~(\ref{eq:twoLorentzians}) we find
\begin{eqnarray}
\Delta \tilde{E}_0&\approx & 2d_{\veczero}V^2 \int_{-C/V^2}^0\!\rmd x
\frac{x+\bar{t}-\pi\tilde{s}_{\vecR}d_{\vecR}}{(x+\bar{t}-\pi\tilde{s}_{\vecR}d_{\vecR})^2
+(\pi d_{\veczero})^2}
\nonumber \\
&&+
2d_{\veczero}V^2 \int_{-C/V^2}^0\!\!\rmd x
\frac{x-\bar{t}+\pi\tilde{s}_{\vecR}d_{\vecR}}{(x-\bar{t}+\pi\tilde{s}_{\vecR}d_{\vecR})^2
+(\pi d_{\veczero})^2}
\nonumber \\
&& - 2d_{\veczero}V^2 
\int_{-\infty}^0\!\rmd x
\frac{\bar{t}-\pi\tilde{s}_{\vecR}d_{\vecR}}{(x+\bar{t}-\pi\tilde{s}_{\vecR}d_{\vecR})^2
+(\pi d_{\veczero})^2}
\nonumber\\
&&+2d_{\veczero}V^2 \int_{-\infty}^0\!\rmd x
\frac{\bar{t}-\pi\tilde{s}_{\vecR}d_{\vecR}}{(x-\bar{t}+\pi\tilde{s}_{\vecR}d_{\vecR})^2
+(\pi d_{\veczero})^2}
\nonumber \\
&=& 4V^2d_{\veczero}\ln\left(\frac{V^2}{C}\right)\nonumber\\
&& +2V^2d_{\veczero}\ln\left[(\bar{t}-\pi\tilde{s}_{\vecR}d_{\vecR})^2
+(\pi d_{\veczero})^2\right]\\
&& - 4V^2d_{\veczero} \frac{\bar{t}-\pi\tilde{s}_{\vecR}d_{\vecR}}{\pi d_{\veczero}} 
\arctan\left( \frac{\bar{t}-\pi\tilde{s}_{\vecR}d_{\vecR}}{\pi d_{\veczero}}\right)
\nonumber
\end{eqnarray}
with $C={\cal O}(1)$.
With the help of eq.~(\ref{eq:noftbarrelationB}) we find
\begin{eqnarray} 
\Delta \tilde{E}_0&=&4 d_{\veczero}V^2
 \ln\left(\pi V^2 d_{\veczero}/C\right) \nonumber\\
&& -4 d_{\veczero}V^2 \ln\left[\cos(\pi/2-n_{1,\uparrow}\pi)\right]
\nonumber\\
&& -4 d_{\veczero}V^2(\pi/2-n_{1,\uparrow}\pi)\tan\left(\pi/2-n_{1,\uparrow}\pi\right) 
\nonumber\\
&& +{\cal  O}\left(V^4\ln(V^2)\right)  \; .
\label{eq:gsenergyBlattice}
\end{eqnarray}

\subsubsection{$\vecR \in \hbox{$A$-lattice}$}

We have to leading order
\begin{equation}
\Delta \tilde{E}_0
\approx 2V^2 \int_{-C/V^2}^0 \rmd x \sum_b xV^2D_b(x)\; .
\end{equation}
Using the expressions~(\ref{eq:twopeaksA})  we find
\begin{eqnarray}
\frac{\Delta \tilde{E}_0}{V^2d_{\veczero}}&=&4  \int_{-C/V^2}^0\rmd x 
\frac{x(x^2+\bar{t}^2(1+p_{\vecR}))}{(x^2-\bar{t}^2(1+p_{\vecR}))^2
+(2\pi d_{\veczero})^2x^2}
\nonumber\\
&=& 4 \int_{-C/(V^2\bar{t}\sqrt{1+p_{\vecR}})}^0
\rmd y \frac{y(y^2+1)}{(y^2-1)^2+\tilde{\kappa}_{\vecR}^2y^2}\nonumber \\
&=&- \int_0^{C^2/(V^2\bar{t}\sqrt{1+p_{\vecR}})^2}
\!\rmd \lambda 
\frac{2\lambda-2+\tilde{\kappa}_{\vecR}^2+4-\tilde{\kappa}_{\vecR}^2}{
\lambda^2-2\lambda+1+\tilde{\kappa}_{\vecR}^2\lambda}\nonumber\\
&=& - \ln\left[\lambda^2-2\lambda+1+\tilde{\kappa}_{\vecR}^2\lambda
\right]_0^{C^2/(V^2\bar{t}\sqrt{1+p_{\vecR}})^2}\nonumber \\
&& - \frac{(4-\tilde{\kappa}_{\vecR}^2)}{\tilde{\kappa}_{\vecR}} G(\tilde{\kappa}_{\vecR})
\nonumber \\
&=& -\left[
4\ln \left(\frac{C\tilde{\kappa}_{\vecR}}{2\pi d_{\veczero}V^2}\right)
+ \frac{(4-\tilde{\kappa}_{\vecR}^2)}{\tilde{\kappa}_{\vecR}} G(\tilde{\kappa}_{\vecR})
\right]
\label{eq:gsenergyAlattice}
\end{eqnarray}
with $p_{\vecR}$ from eq.~(\ref{eq:defalpha})
and $\tilde{\kappa}_{\vecR}$ from eq.~(\ref{eq:defkappatilde}),
and we used eq.~(\ref{eq:tofnAlattice}) in the third step.

\subsection{Level splitting via the RKKY interaction}
\label{sec:levelRKKYsplitting}

In the absence of an electron transfer between the impurity sites,
$t_{12}=0$, the effective level splitting is not necessarily zero.

\subsubsection{$\vecR \in \hbox{$B$-lattice}$}

{}From eq.~(\ref{eq:twoLorentzians}) we see that the impurity
contribution to the density of states is dominantly given by
two Lorentz peaks at $\omega_{\pm}=\pm V^2\pi d_{\vecR}$ with 
$\pi d_{\veczero}V^2$ as half-width at half maximum,
\begin{eqnarray}
V^2D_{1,\sigma}(x) &=& 
\frac{d_{\veczero}}{(x-\pi\tilde{s}_{\vecR}d_{\vecR})^2
+(\pi d_{\veczero})^2}
\; , \nonumber \\
V^2D_{2,\sigma}(x) &=& 
\frac{d_{\veczero}}{(x+\pi\tilde{s}_{\vecR}d_{\vecR})^2
+(\pi d_{\veczero})^2}\; .
\label{eq:twoLorentzianstbarzero}
\end{eqnarray}
Moreover, the particle density in the level $b=1$ is
$n_{1,\uparrow}\approx 1/2-\tilde{s}_{\vecR}d_{\vecR}/(\pi d_{\veczero})\neq 1/2$.

\subsubsection{$\vecR \in \hbox{$A$-lattice}$}

\begin{figure}[t]%
\includegraphics[width=\linewidth]{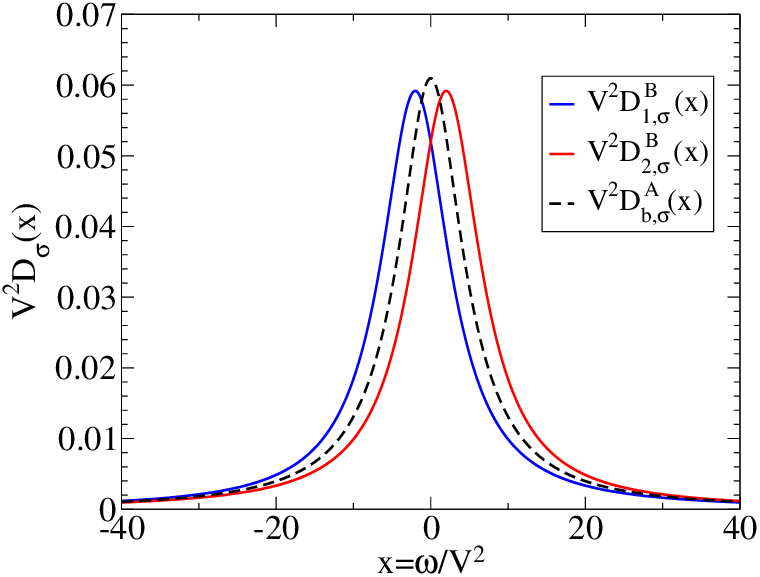}
\caption{Impurity density of states, 
eqs.~(\ref{eq:twoLorentzianstbarzero}) and~(\ref{eq:oneLorentztbarzero}),
as a function of frequency $x=\omega/V^2$
for two impurities at nearest-neighbor distance 
$\vecR=(1,0,0)\in \hbox{$B$-lattice}$ (full lines)
and next-nearest-neighbor distance   
$\vecR=(2,0,0)\in \hbox{$A$-lattice}$ (dashed line)
for $\overline{t}=0$.
Numerically, $d_{\veczero}=1.712$, $\pi d_{(1,0,0)}=2$,
$\tilde{s}_{(1,0,0)}=-1$, and $d_{(2,0,0)}=0.2928$ so that
$\beta_{(2,0,0)}^2=28.08$ and 
$\gamma_{(2,0,0)}^2=1.03$.\label{fig:dosAB}}
\end{figure}

For the $A$-lattice, the situation is different. 
We have 
$D_{1,\sigma}(\omega)=D_{2,\sigma}(\omega)$ and find
a single peak at $\omega=0$.
In particular, for $\vecR=\veczero$ we observe a Lorentz peak at $\omega=0$
with half width at half maximum of $2\pi d_{\veczero}V^2$.
For $\vecR\neq \veczero$ we have
\begin{equation}
V^2 D_{b,\sigma}(x)= \frac{d_{\veczero}}{\beta_{\vecR}^2}
\frac{(x/\beta_{\vecR})^2+1}{
[(x/\beta_{\vecR})^2-1]^2+4\gamma_{\vecR}^2(x/\beta_{\vecR})^2}
\label{eq:oneLorentztbarzero}
\end{equation}
with
\begin{equation}
\beta_{\vecR}^2=\pi^2(d_{\veczero}^2-d_{\vecR}^2)\; , \;
\gamma_{\vecR}^2=\frac{d_{\veczero}^2}{d_{\veczero}^2-d_{\vecR}^2} \;.
\end{equation}
In general, $D_b(x)$ is almost a Lorentz peak at $x=0$ with
half width half maximum $\beta_{\vecR}$ because $\gamma_{\vecR}\approx 1$ 
for all $\vecR\neq \veczero$.
Apparently, the RKKY splitting is absent for $\vecR\in \hbox{$A$-lattice}$.
We only have a single peak in the density of state for $t_{12}=0$
and, correspondingly, eq.~(\ref{eq:tofnAlattice}) gives
$n_{1,\uparrow}=1/2$.
In Fig.~\ref{fig:dosAB}
we show the density of states for two impurities at nearest-neighbor positions
in the simple-cubic lattice, $\vecR=(1,0,0) \in \hbox{$B$-lattice}$, 
and at next-nearest-neighbor positions, $\vecR=(2,0,0) \in \hbox{$A$-lattice}$.

Note that, for the parameters chosen,
the respective sums $\sum_b D_{b,\sigma}(\omega)$ 
are almost indistinguishable. Therefore, it is difficult to resolve
the two-peak structure for $U=0$.
For finite $U$, and for $U\gg 1$ in particular,
the peaks narrow into two well-separated, sharp Kondo resonances
whose splitting can be detected more easily.

\subsection{RKKY energy}
\label{sec:RKKYenegry}

Lastly, we discuss the contribution of the two-impurity interaction
to the ground-state energy for $t_{12}=0$ and $V\ll 1$.

\begin{figure}[t]%
\includegraphics[width=\linewidth]{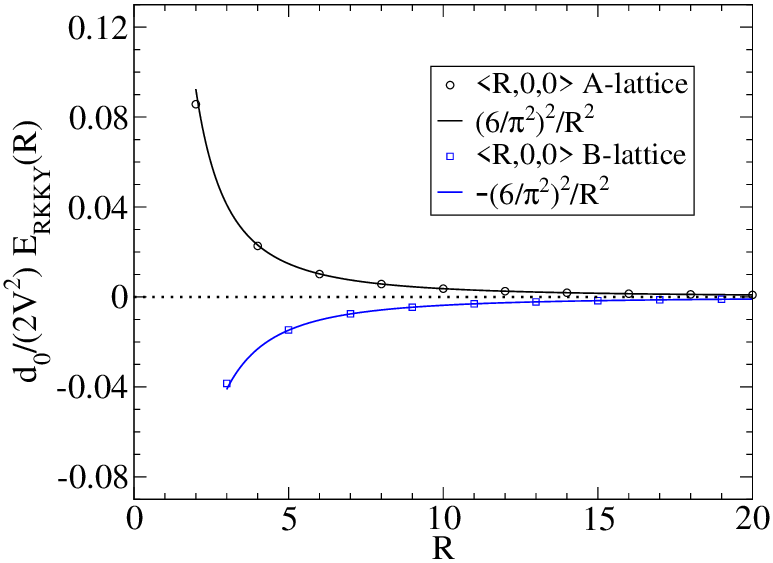}
\includegraphics[width=\linewidth]{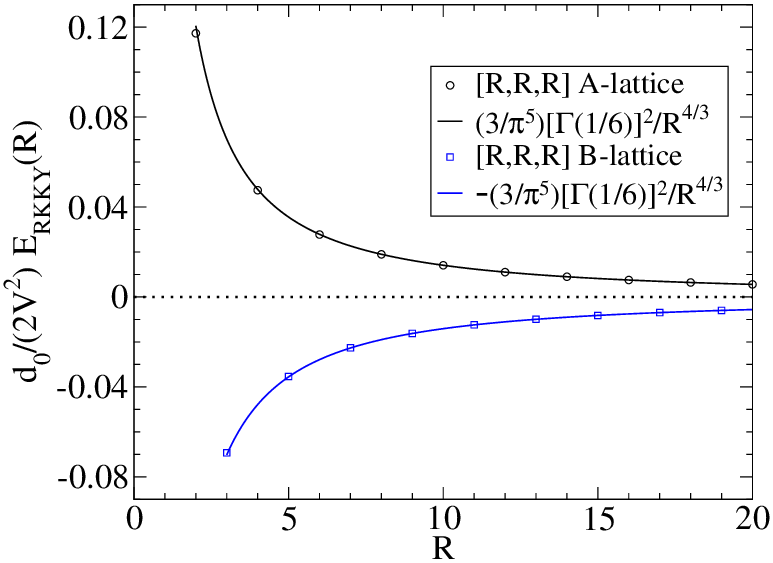}
\caption{RKKY energy, eqs.~(\ref{eq:RKKYBlattice}) and~(\ref{eq:RKKYAlattice}),
as a function of the impurity  
separation~$R$ along the axis directions $<\!\!R,0,0\!\!>$ (top)
and along the diagonal direction $[R,R,R]$ (bottom).\label{fig:RKKYenergy}}
\end{figure}

\subsubsection{$\vecR \in \hbox{$B$-lattice}$}

The level splitting results in a decrease in energy.
{}From eq.~(\ref{eq:gsenergyBlattice}) we see that for $y=n_{1,\uparrow}\pi-\pi/2$
and with $\Delta E_0=\Delta \tilde{E}_0$ for $t_{12}=0$
\begin{eqnarray}
\Delta E_0&=&
4 d_{\veczero}V^2 \ln\left(\pi V^2 d_{\veczero}/C\right) \nonumber\\
&&  -4 d_{\veczero}V^2\left[\ln\left(\cos(y)\right)+y \tan(y) \right]\; .
\end{eqnarray}
As seen from eq.~(\ref{eq:noftbarrelationB}), 
the absolute value of $\tan(y)=-\tilde{s}_{\vecR}d_{\vecR}/d_{\veczero}$ is small so that 
\begin{equation}
\Delta E_0\approx 
4 d_{\veczero}V^2 \ln\left(\pi V^2 d_{\veczero}/C\right) 
-2 d_{\veczero}V^2\left(\frac{d_{\vecR}}{d_{\veczero}}\right)^2\; .
\end{equation}
The first term originates from the host electron scattering off the individual impurities,
the second term reflects the energy gain by the coherent scattering off
both impurity levels. Therefore, 
the RKKY energy of the two impurities is given by
\begin{equation}
E_{\rm RKKY}(\vecR\in \hbox{$B$-lattice})
= -2d_{\veczero}V^2 \left(\frac{d_{\vecR}}{d_{\veczero}}\right)^2\; .
\label{eq:RKKYBlattice}
\end{equation}
The RKKY interaction always leads to a {\sl decrease\/} in energy when
$\vecR \in \hbox{$B$-lattice}$.
We show the RKKY energy in Fig.~\ref{fig:RKKYenergy}.

\subsubsection{$\vecR \in \hbox{$A$-lattice}$}

The impurity levels are not split 
when $\vecR \in \hbox{$A$-lattice}$. Nevertheless, 
the coherent scattering off the two impurities changes the ground-state energy. 
With $\tilde{\kappa}_{\vecR}=2/\sqrt{1-(d_{\vecR}/d_{\veczero})^2}\gtrsim 2$
we obtain from eq.~(\ref{eq:gsenergyAlattice})
\begin{equation}
\Delta E_0\approx 4d_{\veczero}V^2\ln(\pi V^2d_{\veczero}/C)
+2V^2d_{\veczero}\left(\frac{d_{\vecR}}{d_{\veczero}}\right)^2\; .
\end{equation}
The functional dependence on $\vecR$ is the same as in~(\ref{eq:RKKYBlattice}),
\begin{equation}
E_{\rm RKKY}(\vecR\in \hbox{$A$-lattice})
= 2d_{\veczero}V^2 \left(\frac{d_{\vecR}}{d_{\veczero}}\right)^2 \; , 
\label{eq:RKKYAlattice}
\end{equation}
but for $\vecR \in \hbox{$A$-lattice}$
the RKKY interaction always leads to an {\sl increase\/} in energy,
see Fig.~\ref{fig:RKKYenergy}.
The functional form of the RKKY energy can also be obtained from 
the RKKY exchange interaction, as we show in appendix~\ref{app:RKKYexchange}.

\section{Conclusions}
\label{sec:conclusions}

In this work we studied the non-interacting two-impurity Anderson model on a lattice.
We calculated the single-particle Green functions analytically 
using the equation-of-motion method. We focused on the particle-hole
symmetric case at half band-filling for host electrons with nearest-neighbor
electron transfer on a simple-cubic (bipartite) lattice. 
For a local, weak hybridization between impurities and host electrons
and no direct coupling between
the impurities (RKKY limit), we provide explicit analytic formulae for
the impurity density of states and the ground-state energy
contribution due to the presence of the impurities.

In general, the RKKY interaction leads to an effective coupling between the
impurities so that the single-particle density of states displays two peaks.
For the non-interacting Anderson model, the level splitting is
small and decays as a function of the impurity distance.
However, even for short distances, 
two peaks are difficult to resolve in the total density of states.
The splitting exactly vanishes only if the host electrons' 
density-density correlation function is zero, as is the case 
when both impurities are on the same sublattice, due to particle-hole symmetry
at half band-filling.

The analytic formulae for the RKKY interaction in the 
non-interacting two-impurity Anderson model differ
even to second-order perturbation theory in the hybridization
because the exact solution contains all multiple scatterings between
impurities and host electrons. Nevertheless, the perturbative approach
is able to account for the distance-dependence qualitatively because
it reproduces the alternating sign between the two sublattices.
In general, however, the multiple scattering contributions modify the exchange
interaction quantitatively for a dense host-electron system, namely in its size and
distance-dependence.

The  two-impurity Anderson model at $U=0$ contains the RKKY indirect exchange
mechanism but it does not capture the
local-moment formation on the impurity sites that is present in the
Kondo limit of half band-filling and large interaction strengths, $U\gg W$.
However, the competition of Kondo and RKKY physics can be studied variationally 
using the Gutzwiller approach that starts from a single-particle product state.
Therefore, the present formulae are an indispensable prerequisite for
the analytic evaluation of Gutzwiller-correlated wave 
functions. Work in this direction is in progress~\cite{GutzwillerTIAM}.


\begin{acknowledgement}
Z.M.M.\ Mahmoud thanks the Fachbereich Physik at the Philipps Universit\"at Marburg
for its hospitality.
\end{acknowledgement}

\appendix

\section{Particle-hole transformation}
\label{app:phop}

In this appendix we derive the unitary operator that generates the particle-hole
transformation~$\tau_{\rm ph}$ in eq.~(\ref{eq:phtrafodef}). 
For the host electrons we use 
$\hat{\tau}_{c;\sigma}=\prod_{\vecR} \hat{\tau}_{\vecR;\sigma}^{\vphantom{+}}$,
where 
\begin{equation}
\hat{\tau}_{\vecR;\sigma}^{\vphantom{+}}=
- \hat{c}_{\vecR,\sigma}^++e^{-\rmi \vecQ\cdot \vecR}\hat{c}_{\vecR,\sigma}^{\vphantom{+}}
\end{equation}
acts only on site $\vecR$ of the lattice.
We have $\hat{\tau}_{\vecR;\sigma}^{\vphantom{+}}\hat{\tau}_{\vecR;\sigma}^+ 
= \openone$ and
\begin{equation}
\hat{\tau}_{\vecR;\sigma}^+\hat{c}_{\vecR,\sigma}^{\vphantom{+}}
 \hat{\tau}_{\vecR;\sigma}^{\vphantom{+}}=
-e^{\rmi \vecQ\cdot \vecR} \hat{c}_{\vecR,\sigma}^+ 
\; .
\end{equation}
Note that the operators $ \hat{\tau}_{\vecR;\sigma}^{(+)}$ 
and $ \hat{\tau}_{\vecR^{\prime};\sigma}^{(+)}$ anti-commute
with each other and with $\hat{c}_{\vecR^{\prime\prime}}^{(+)}$
for $\vecR\neq \vecR^{\prime}$ and $\vecR^{\prime\prime}\neq \vecR,\vecR^{\prime}$.
Consequently, the unitary operator
$\hat{\tau}_{c;\sigma}$ generates the par\-ticle-hole transformation 
for the host electrons ($L$ is even),
\begin{eqnarray}
\hat{\tau}_{c;\sigma}^+\hat{c}_{\veck,\sigma}^{\vphantom{+}}
 \hat{\tau}_{c;\sigma}^{\vphantom{+}}&=&
(-1)^{L-1}\sqrt{\frac{1}{L}} \sum_{\vecR} e^{-\rmi \veck \cdot \vecR}
\hat{\tau}_{\vecR;\sigma}^+\hat{c}_{\vecR,\sigma}^{\vphantom{+}}
 \hat{\tau}_{\vecR;\sigma}^{\vphantom{+}}\nonumber \\
&=&
\sqrt{\frac{1}{L}} \sum_{\vecR} e^{-\rmi \veck \cdot \vecR}e^{\rmi \vecQ \cdot \vecR}
\hat{c}_{\vecR,\sigma}^+ = \hat{c}_{\vecQ-\veck,\sigma}^+
\; .\nonumber \\
\end{eqnarray}
Furthermore, for the impurity electrons we set
\begin{equation}
\hat{\tau}_{b;\sigma}^{\vphantom{+}}=
\hat{d}_{b,\sigma}^++e^{-\rmi \vecQ\cdot \vecR_b}\hat{d}_{b,\sigma}^{\vphantom{+}}
\end{equation}
with $\hat{\tau}_{b;\sigma}^{\vphantom{+}}\hat{\tau}_{b;\sigma}^+=\openone$
and find
\begin{equation}
\hat{\tau}_{2;\sigma}^+\hat{\tau}_{1;\sigma}^+
\hat{d}_{b,\sigma}^{\vphantom{+}}
 \hat{\tau}_{1;\sigma}^{\vphantom{+}} \hat{\tau}_{2;\sigma}^{\vphantom{+}}
= -e^{\rmi \vecQ\cdot \vecR_b} \hat{d}_{b,\sigma}^+ 
\end{equation}
because $\hat{\tau}_{2;\sigma}^{(+)}$ and $\hat{\tau}_{1;\sigma}^{(+)}$
anti-commute.
The unitary operator for the complete particle-hole transformation reads
\begin{equation}
\hat{\tau}_{\rm ph}^{\vphantom{+}}
=\prod_{\sigma} \prod_{\vecR} \hat{\tau}_{\vecR;\sigma}^{\vphantom{+}}
\prod_b \hat{\tau}_{b;\sigma}^{\vphantom{+}} \; .
\end{equation}

\section{RKKY interaction at large distances}
In this appendix we derive the asymptotic formulae applied in Fig.~\ref{fig:RKKYenergy}.

\subsection{Axis direction}

We must calculate
\begin{equation}
d_{(R,0,0)}=\frac{6}{\pi}\int_0^{\infty} \rmd x J_R(x)[J_0(x)]^2 
\end{equation}
for $R\gg 1$. Now that $J_R(x)\sim x^R$ for small $x$,
we can safely use the asymptotic expression for 
$J_0(x\gg1)$, number~9.2.1 of~\cite{abramowitzstegun}, 
to arrive at the approximation
\begin{equation}
d_{(R,0,0)}\approx \frac{6}{\pi}\int_0^{\infty} \rmd x J_R(x) \frac{2}{\pi x}\cos^2(x-\pi/4)
\; .
\end{equation}
{\sc Mathematica}~\cite{Mathematica} gives an analytic result for this integral.
Ignoring an exponentially small oscillating term we find
\begin{equation}
d_{(R,0,0)}\approx \frac{6}{\pi^2}\frac{1}{R} \; .
\end{equation}
As seen from the figure, this approximate formula works very well for all
$R\geq 3$.

\subsection{Diagonal direction}

We must calculate for $\vecR=(R,R,R)$
\begin{equation}
d_{\vecR}=\frac{6R}{\pi}\int_0^{\infty} \rmd y [J_R(Ry)]^3 
\end{equation}
for $R\gg 1$. We use number~9.3.6 together with numbers~9.3.38/39
of~\cite{abramowitzstegun}
to arrive at
\begin{eqnarray}
d_{\vecR}&\approx& 
\frac{6R}{\pi}\int_0^1 \rmd y 
\left(\frac{4 \zeta_1(y)}{1-y^2}
\right)^{3/4} 
\left[{\rm Ai}\left(R^{2/3}\zeta_1(y)\right)\right]^3 \nonumber \\
&& +
\frac{6R}{\pi}\int_1^{\infty} \rmd y 
\left(\frac{4 \zeta_2(y)}{1-y^2}
\right)^{3/4} 
\left[{\rm Ai}\left(R^{2/3}\zeta_2(y)\right)\right]^3 \; ,\nonumber \\
\end{eqnarray}
where ${\rm Ai}(x)$ is the first Airy function~\cite{abramowitzstegun} and
\begin{eqnarray}
\frac{2}{3} [\zeta_1(y)]^{3/2}&=& \ln\left(\frac{1+\sqrt{1-y^2}}{y}\right)
-\sqrt{1-y^2} \; ,\nonumber \\
\frac{2}{3} [-\zeta_2(y)]^{3/2}&=& \sqrt{y^2-1}-\arccos(1/y) \; .
\end{eqnarray}
Since ${\rm Ai}(|x|\to \infty)=0$, $\zeta_{1,2}(y)$ must be small.
Therefore, $|y-1|\ll 1$ and
$\zeta_{1,2}\approx 2^{1/3}(1-y)$.
With the approximation 
$[4\zeta_{1,2}(y)/(1-y^2)]^{3/4}\approx 2$ for the relevant integration region
we can simplify
\begin{eqnarray}
d_{\vecR}&\approx& \frac{12}{\pi}\int_0^{\infty} \rmd y
\left[{\rm Ai}\left(R^{2/3}2^{1/3}(1-y)\right)\right]^3\nonumber \\
&\approx& R^{-2/3} \frac{12}{\pi}2^{-1/3}\int_{-\infty}^{\infty}\rmd \lambda
\left[{\rm Ai}(\lambda)\right]^3\; .
\label{eq:almostthere}
\end{eqnarray}
With the Fourier representation of the Airy function,
\begin{equation}
{\rm Ai}(\lambda)=\int_{-\infty}^{\infty}\frac{\rmd s}{2\pi} 
e^{\rmi s\lambda} e^{-\rmi s^3/3} \; , 
\end{equation}
we find
\begin{equation}
\int_{-\infty}^{\infty}\rmd \lambda
\left[{\rm Ai}(\lambda)\right]^3=
\iint\limits_{-\infty}^{\hphantom{++}+\infty}
\frac{\rmd s_1}{2\pi} 
\frac{\rmd s_2}{2\pi}
e^{\rmi (s_1^3+s_2^3-(s_1+s_2)^3)/3} \; .
\end{equation}
Using $s_1=(u+v)/\sqrt{2}$ and $s_2=(u-v)/\sqrt{2}$ this can be written as
\begin{equation}
\int_{-\infty}^{\infty}
\rmd \lambda
\left[{\rm Ai}(\lambda)\right]^3=
\int_{-\infty}^{\infty}\frac{\rmd u}{2\pi} 
e^{\rmi u^3/\sqrt{2}}
\int_{-\infty}^{\infty}\frac{\rmd v}{2\pi}
e^{-\rmi uv^2/\sqrt{2}} \; .
\end{equation}
The remaining two integrals can be carried out analytically using 
{\sc Mathematica}~\cite{Mathematica} to give
\begin{equation}
\int_{-\infty}^{\infty}\rmd \lambda
\left[{\rm Ai}(\lambda)\right]^3=
\frac{2^{1/3}\sqrt{\pi}}{4\sqrt{3}\pi^2}\Gamma\left(\frac{1}{6}\right)\; ,
\end{equation}
where $\Gamma(x)$ is the Gamma function.
We insert this result into eq.~(\ref{eq:almostthere})
and finally find
\begin{equation}
d_{(R,R,R)}= \sqrt{3}\pi^{-5/2}\Gamma\left(\frac{1}{6}\right) R^{-2/3}
\;.
\end{equation}
The prefactor evaluates to 
$\sqrt{3}\pi^{-5/2}\Gamma(1/6) \approx 0.55113$.

\section{RKKY exchange interaction}
\label{app:RKKYexchange}

Second-order perturbation theory in the impurity-host hybridization
gives the RKKY indirect exchange interaction between two spins on a lattice
in a metallic host, see~\cite{Solyom}, appendix~I,
\begin{equation}
\hat{H}_{\rm RKKY}=-\sum_{\vecR,\vecR'}J(\vecR-\vecR')
\hat{\vecS}_{\vecR}\cdot\hat{\vecS}_{\vecR'}
\end{equation}
with the exchange interaction between two spins at distance $\vecR$
\begin{equation}
J(\vecR)= \left(\frac{V}{L}\right)^2 \sum_{\veck,\vecp}
\frac{n_{\veck,\uparrow}^{\rm FS}(1-n_{\vecp,\uparrow}^{\rm FS})}{
\epsilon(\vecp)-\epsilon(\veck)}
e^{-\rmi (\veck-\vecp)\cdot \vecR} \; .
\end{equation}
With $\vecp'=\vecQ-\vecp$ and with eqs.~(\ref{eq:defDALamA}) 
and~(\ref{eq:defDBLamB}) we can cast the exchange interaction into the form
\begin{eqnarray}
J(\vecR)&=& -(-1)^\vecR 
\int_{-\infty}^0\rmd \epsilon_1
\int_{-\infty}^0\rmd \epsilon_2 
\frac{V^2}{\epsilon_1+\epsilon_2}
\nonumber\\
&& \hphantom{-(-1)^\vecR \int_{-\infty}^0}
D_{A/B}(\epsilon_1,\vecR)D_{A/B}(\epsilon_2,\vecR)\; .
\label{eq:exchangePT}
\end{eqnarray}
Apparently, the RKKY exchange interaction changes its sign when we go
from the $A$-lattice to the $B$-lattice, as it should.

Eq.~(\ref{eq:exchangePT}) gives the qualitatively correct dependence on~$\vecR$
only when $\vecR\in\hbox{$A$-lattice}$, namely
$J(\vecR)\propto d_{\vecR}^2$. For 
$\vecR\in\hbox{$B$-lattice}$, second-order perturbation theory is not accurate enough
because it does not account for the level splitting shown in Fig.~\ref{fig:dosAB}.
Therefore, it does not reproduce the correct behavior 
$J(\vecR)\propto d_{\vecR}^2$ for 
$\vecR\in\hbox{$B$-lattice}$.


\providecommand{\WileyBibTextsc}{}
\let\textsc\WileyBibTextsc
\providecommand{\othercit}{}
\providecommand{\jr}[1]{#1}
\providecommand{\etal}{~et~al.}

\end{document}